\documentclass[11pt]{article}
\usepackage[english]{babel}
\usepackage{graphics}
\usepackage{graphicx}
\usepackage{xcolor}
\usepackage{amssymb}
\usepackage{amsmath}
\usepackage{amsthm}
\usepackage{amsfonts}
\usepackage{braket}
\usepackage{xcolor}
\usepackage{authblk}
\usepackage{dutchcal}
\usepackage{tikz,pgfplots}

\usepackage[pdfstartview=XYZ,
bookmarks=true,
colorlinks=true,
linkcolor=blue,
urlcolor=blue,
citecolor=blue,
pdftex,
bookmarks=true,
linktocpage=true, 
hyperindex=true
]{hyperref}

\usepackage{orcidlink}

\theoremstyle{plain}
\newtheorem{theorem}{Theorem}[section]
\newtheorem{lemma}[theorem]{Lemma}

\theoremstyle{definition}
\newtheorem{definition}[theorem]{Definition}

\theoremstyle{remark}
\newtheorem{remark}{Remark}

\begin{document}

\title{Quantum Hitting Time according to a given distribution}

\author[1]{Paola Boito \orcidlink{0000-0002-3559-393X} }
\author[2]{G.~M. Del Corso\thanks{Corresponding author: G.~M. Del Corso. Email: gianna.delcorso@unipi.it} \orcidlink{0000-0002-5651-9368}}

\affil[1]{Dipartimento di Matematica, Universit\`a di Pisa, Largo B. Pontecorvo 5, 56127 Pisa, Italy.}
\affil[2]{Dipartimento di Informatica, Universit\`a di Pisa, Largo B. Pontecorvo 3, 56127 Pisa, Italy.}

\providecommand{\keywords}[1]{\textbf{\textit{Index terms---}} #1}

\date{}

\maketitle


\begin{abstract}
In this work we focus on the notion of quantum hitting time for discrete-time Szegedy quantum walks, compared to its classical counterpart. Under suitable hypotheses, quantum hitting time is known to be of the order of the square root of classical hitting time: this quadratic speedup is a remarkable example of the computational advantages associated with quantum approaches. 

Our purpose here is twofold. On one hand, we provide a detailed proof of quadratic speedup for time-reversible walks within the Szegedy framework, in a language that should be familiar to the linear algebra community. 
Moreover, we explore the use of a general distribution in place of the stationary distribution in the definition of quantum hitting time, through theoretical considerations and numerical experiments.
\end{abstract}

\begin{keywords}
Quantum walks; quantum hitting time; search on graphs
\end{keywords}

\section{Introduction}
The hitting time of a random walk on a graph ${G}$ is the expected number of steps required to reach a certain node starting from a given node or a given distribution. A crucial application is in the search problem, where hitting time tells us how many steps are needed to detect a marked node. Hitting time has a role in the analysis of complex networks, where it represents a measure of communicability for the underlying graph \cite{xia2019random}. More applications include the link prediction problem \cite{liben2003link} and clustering \cite{chen2008clustering}; see also
\cite{fasino2021hitting} for an analysis of hitting time of second-order random walks.

The doubly averaged hitting time with respect to the stationary distribution $\pi$ of a random walk
$$ K({G})=\sum_{i=1}^n\sum_{j=1}^n \pi_i \pi_j H_{ij}$$
coincides with the {\em Kemeny constant} of ${G}$ \cite{bini2018kemeny, kirkland2021directed}, which has been used to provide centrality measures for graphs (see e.g. \cite{altafini2022centrality} for a recent application).
Also note that Kemeny's constant is closely related to the graph resistance or Kirchhoff index of a graph \cite{palacios2010kirchhoff, wang2017kemeny}, which plays a role as well in network analysis \cite{ghosh2008minimizing, klein2010centrality, estrada2010vibrational, ellens2011effective}. See \cite{apers2022elfs} for a recent perspective on random and quantum walks, electric hitting time and graph resistance.

In a quantum framework, random walks are replaced by quantum walks, which exhibit peculiar properties; see for instance the reviews \cite{venegas2012quantum, kadian2021quantum} or the book \cite{PortugalBook}. In particular, quantum walks typically tend to diffuse faster on a graph than classical random walks. One way in which this remark can be made more precise is through the definition of a quantum notion of hitting time. We focus here on quantum hitting time for discrete-time quantum walks. Definitions and applications of hitting time have also been formulated for continuous-time quantum walks: see e.g., \cite{Bai2019, Emms2008, Varbanov2008, roland2018finding}. 
However, we will not make use of the continuous-time formulation in this work.

Historically, the interest for quantum hitting time arises mainly as an estimate of the running time for quantum search algorithms. We mention in passing that the well-known Grover search algorithm can be recast as a quantum walk over a complete graph. For symmetric Markov chains, it has been shown \cite{Szegedy2004} that the quantum hitting time exhibits a quadratic speedup with respect to its classical counterpart. This seminal result has later been generalized to nonsymmetric walks, with extensions to the problem of finding marked elements rather than merely detecting their presence \cite{magniez2007search, krovi2010adiabatic, magniez2012hitting, krovi2016quantum}. Recent progress in this direction has also been made for continuous-time quantum walks \cite{chakraborty2020finding, apers2022quadratic}. 
 
Quadratic speedup has been shown to be optimal under reversibility hypotheses \cite{magniez2009proceedings}. Note that, in general, one cannot hope for a quadratic speedup in quantum hitting time for nonreversible Markov chains. As a counterexample, consider an $n$-node directed cycle with self-loops \cite{krovi2010adiabatic}. The associated Markov chain is ergodic but not reversible; classical hitting time is of order $\Theta(n)$ and any quantum operator acting locally on this cycle needs time $\Omega(n)$ to find a marked vertex. 

Like classical hitting time, quantum hitting time has also been applied to network analysis: in \cite{bai2019quantum}, for instance, the authors use commute time in the context of similarity measures for complete weighted graphs. It is unclear, though, if and how one may define a quantum equivalent of Kemeny's constant. 

More generally, limiting distributions of quantum walks can be used to provide centrality measures for networks, much like in the classical case; a popular example is the Quantum Page Rank algorithm by Paparo et al. \cite{paparo2013quantum}.

Several authors have observed that quantum walks may highlight properties of the underlying graph that go undetected by classical random walks. This provides additional motivation for the use of quantum hitting time in the analysis of graphs and networks.

The definition of quantum hitting time \cite{magniez2011search} requires knowledge of the classical stationary distribution $\pi$ associated with the given graph. By analogy with classical hitting time, which can be defined w.r.t.~an arbitrary distribution, 
we investigate an extension of quantum hitting time involving a general distribution rather than the stationary one. Motivation comes from the fact that, in some cases, computation of $\pi$ may be a computationally intensive process. It could be advantageous to replace $\pi$ by a (more or less accurate) approximation. Stationary distributions are typically computed via an iterative method (e.g., the power method), so one might apply just a few iterative steps to obtain a rough approximation of $\pi$ and then use that approximation in the computation of quantum hitting time. It is also interesting -- again, analogously to the classical case -- to investigate the behavior of quantum hitting time according to a distribution concentrated in a single node.

The present work is also an effort to make quantum walks better known by the numerical linear algebra community: for this reason we have tried to be as self-contained as possible and we have made a point of spelling out in detail the proof of quantum quadratic speedup (Theorem \ref{thm:quadspeedup}), so as to give a view of typical quantum walk techniques in the language of linear algebra. We assume however that the reader has a basic knowledge of the main principles of quantum mechanics and of the bra-ket notation.

We recall definitions and properties of classical and quantum walks on graphs in section \ref{sec:background}. Quantum hitting time is presented in section \ref{sec:quantumht} for the symmetric and the time-reversible case. In Section \ref{sec:speedup} we present a detailed proof of quadratic speedup for quantum hitting time for time-reversible walks, by extending Szegedy's original approach.
Section \ref{sec:generalized} is devoted to the extension of quantum hitting time to general distributions, whereas Section \ref{sec:experiments} reports the results of numerical experiments on several examples of graphs, including cycles and synthetic scale-free graphs. Finally, section \ref{sec:conclusions} draws conclusions and outlines possible future developments.

\section{Background}\label{sec:background}

In this section we recall a few  basic notions about discrete-time classical random walks and quantum walks, and fix notation that will be used in the following. Consider a connected graph ${G}=(V,E)$, where $V$ is the set of nodes or vertices and $E\subset V\times V$ the set of edges, and let $n$ be the number of nodes. Recall that the adjacency matrix of ${G}$ is $A\in\mathbb{R}^{n\times n}$, such that $A_{ij}\geq 0$ denotes the weight of edge $(i,j)$. If the weights take values in $\{0,1\}$, then ${G}$ is an unweighted graph; moreover, ${G}$ is undirected if and only if $A$ is symmetric. 

In the graph search problem, one typically assumes that the nodes in a certain subset $M\subset V$ of cardinality $m$ are {\em marked}, and seeks to detect and/or find them. 
Random walks can be used as a tool to solve this problem with sufficiently high probability. To this end, the notion of {\em hitting time} plays a crucial role, as it tells us how fast a random walk spreads over the graph. Therefore, hitting time helps predict how many walk steps need to be performed in order to find a marked node with high probability.

\subsection{Classical random walks}\label{subsec:CRW}
A classical discrete-time random walk on ${G}$ is a finite-state stochastic process identified by a row-stochastic matrix $P=(P_{ij})\in\mathbb{R}^{n\times n}$, where $P_{ij}$ is the probability of transitioning from node $i$ to node $j$. Clearly such a random walk is a Markov chain, and in fact we will use the two expressions interchangeably.

Probability distributions on our Markov chain will be denoted, as customary, by stochastic row vectors in $\mathbb{R}^n$. Applying one step of the walk to a given distribution $v^T\in\mathbb{R}^n$ amounts to computing the product $v^T P$.
 
Usually, for each index $i$, probabilities $P_{ij}$, $j=1,\ldots,n$ are chosen proportionally to the weights of the edges leaving node $i$:
$$
P_{ij}=\frac{A_{ij}}{\sum_{j\in V}A_{ij}},
$$
that is, $P$ is obtained from $A$ after normalizing its rows w.r.t. the 1-norm. 

Recall that a Markov chain (or, equivalently, its associated transition matrix) is said to be {\em irreducible} if for any pair of nodes $i,j$ there is a nonzero probability of transitioning from node $i$ to node $j$ after some (finite) number of steps; this amounts to saying that the underlying graph must be strongly connected.  An irreducible chain is {\em ergodic} if it is aperiodic, that is, if there exists an integer $k_0\geq 1$ such that for any $k\geq k_0$ all entries of matrix $P^k$ are strictly positive.

If $P$ is irreducible, the Perron-Frobenius theorem ensures the existence of the unique, strictly positive stationary distribution $\pi\in\mathbb{R}^n$ such that $\pi^T P=\pi^T$. In particular, for a random walk on an undirected, unweighted graph it holds $\pi_i=\deg(i)/2N$, where $\deg(i)$ is the degree of node $i$ and $N$ is the number of edges. 

If $P$ is ergodic, all the $n-1$ eigenvalues of $P$ that are different from $1$ have absolute value strictly smaller than $1$. 

Given an irreducible random walk defined by transition matrix $P$, the associated {\em reversed walk} has transition matrix $P^*$ characterized by the balance equation
\begin{equation}
P_{ij}\pi_i=\pi_j P^*_{ji},\label{eq:balance}
\end{equation}
or, equivalently, as $P^*={\rm{diag}}(\pi)^{-1} P^T {\rm{diag}(\pi)},$
where $\pi=(\pi_i)_{i\in V}$ is the stationary distribution of the irreducible Markov chain described by $P$, and such that $\sum_i \pi_i=1$. If  $P=P^*$ then $P$ is called time-reversible. Note that for any undirected graph the associated $P$ is time-reversible. 

The {\em hitting time} $H_{ij}$, which is formally defined below, can be seen as the expected time (i.e., number of steps) needed to visit node $j$, starting from node $i$. It acts as a measure of proximity between nodes $i$ and $j$. A related notion is {\em commute time}: the commute time $C_{ij}$ between nodes $i$ and $j$ is defined as the expected time for a walker to go from node $i$ to node $j$ and back. Clearly it holds $C_{ij}=H_{ij}+H_{ji}$. 

\begin{definition}
The {\em hitting time} $H_{ij}$ from node $i$ to node $j$ is 
$$
H_{ij}=\sum_{t=0}^{\infty} t P_{ij}(t),
$$
where $P_{ij}(t)$ denotes the probability of reaching node $j$ for the first time at time $t$, having left node $i$ at $t=0$. If $i=j$, define $H_{ij}=0$. 
\end{definition}
An equivalent recursive definition of hitting time can be given as
$$
H_{ij}=\left\{\begin{array}{ll}
1+\sum_{k\in\mathcal{N}_i}P_{ik}H_{kj} & {\rm if}\,\, i\neq j,\\
0&{\rm if}\,\, i= j,
\end{array}\right.
$$
where $\mathcal{N}_i$ is the neighbor set of node $i$.

Several generalizations are possible. The target node can be replaced by a set of marked vertices $M\subset V$, in which case one defines
$$H_{iM}=\sum_{t=0}^{\infty} t p_{iM}(t),$$
where $p_{iM}(t)$ is the probability of reaching any node in $M$ for the first time at time $t$, having left node $i$ at $t=0$.
Moreover, the starting node can be chosen from a (normalized) distribution $\sigma$ defined on $V$; this notion of hitting time will be denoted as $H_{\sigma,M}$. It holds
$$
H_{\sigma,M}=\sum_{i\in V}\sigma_i H_{iM}
$$
or, equivalently,
$$
H_{\sigma,M}=\sum_{t=0}^{\infty}p(T\geq t),
$$
where $p(T\geq t)$ is the probability of reaching $M$ for the first time at any time $T\geq t$.

An alternative approach to the notion of hitting time $H_{\sigma,M}$ relies on {\em absorbing walks}. Starting from ${G}$ and from the associated ergodic and reversible Markov chain, define the modified graph $\tilde{G}$ by removing all arcs leaving the marked vertices and replacing them with loops. This means that, once the walk lands on a marked vertex, it cannot leave. The modified transition matrix is $\tilde{P}$ such that
$$
\tilde{P}_{ij}=\left\{\begin{array}{ll}
P_{ij}, & i\notin M \\
\delta_{ij}, & i\in M, 
\end{array}\right.
$$
where $\delta_{ij}=0$ if $i\neq j$ and $\delta_{ij}=1$ if $i=j$.
Up to node relabeling, we can assume that the transition matrix has the form
$$
\Tilde{P}=\left[\begin{array}{cc}
   I_m& 0\\
    * &  P_{-M}
\end{array}
\right],
$$
where $P_{-M}$ is the matrix obtained from $P$ by taking out all rows and columns associated with a node in the marked set, and $I_m$ is the identity matrix of size equal to $m$,  the cardinality of set $M$.

Then the hitting time can be rewritten as~\cite{PortugalBook}
\begin{equation} \label{eq:Hsigma}
H_{\sigma, M}=\sigma_{-M}^T (I-P_{-M})^{-1} {\bf 1}_{-M},
\end{equation}
where $\sigma_{-M}$ is the distribution vector  restricted to the unmarked vertices and ${\bf 1}_{-M}$ denotes the column vector of length $n-m$ with entries all equal to $1$. 
Typically, in this definition, the stationary or the uniform distribution are used.


\subsection{Quantum walks}\label{subsec:QRW}
We adopt Szegedy's approach to the quantization of discrete-time random walks \cite{szegedy2004quantum}. (Another popular approach to discrete-time quantum walks is the {\em coined} formalism, which we do not address here).

Szegedy's definition of quantum walk relies on the notion of graph {\em bipartization}. 
The bipartite graph $\mathcal{G}$ associated with $G$ is defined as follows. The set of vertices is  given by the union of two copies of $V$, which we will call $X$ and $Y$. To each edge $(i,j)\in E$ correspond two edges $(x_i,y_j),(y_i,x_j)\in\mathcal{E}$, where $x_k\in X$ and  $y_k\in Y$ denote the nodes in $X$ and in $Y$, respectively, associated with node $k\in V$, for $k=1,\ldots,n$.

Szegedy's theory was originally developed 
under the hypothesis that $P=P^T$. In this presentation we adopt the generalization to nonsymmetric walks put forward by \cite{magniez2007search}, which relies on the notion of time-reversed walks as defined by the balance equation \eqref{eq:balance}. 

In order to quantize the walk, let us introduce a Hilbert state space with its computational basis:
$$\mathcal{H}={\rm Span}\,\{\ket{x}\ket{y},\,x\in X, y\in Y\}\cong \mathbb{C}^{n^2}.$$
The usual isomorphism between $\mathcal{H}$ and $\mathbb{C}^{n^2}$, where the computational basis of $\mathcal{H}$ is in a one-to-one correspondence with the canonical basis of $\mathbb{C}^{n^2}$, will allow us to switch back and forth between the ket notation for states in $\mathcal{H}$ and the customary linear algebra notation for vectors in $\mathbb{C}^{n^2}$.

Consider the states
$$ \ket{\phi_x}=\ket{x} \otimes \ket{p_x}=\sum_{y\in Y}\sqrt{P_{xy}}\ket{x}\ket{y},\qquad x\in X, $$
where $p_x$ is the row of index $x$ in $P$, and
$$\ket{ \psi_{y}}=\ket{p^*_y}\otimes \ket{y}=\sum_{x\in X}\sqrt{P^*_{yx}}\ket{x}\ket{y},\qquad y\in Y, $$
where $p^*_y$ is the row of index $y$ in $P^*$.

Let 
\begin{equation}
A=(\phi_x)_{x\in X}\label{eq:Adef}
\end{equation} 
be the matrix whose columns are the vectors $\phi_x$, and let \begin{equation}
B=(\psi_y)_{y\in Y}\label{eq:Bdef}
\end{equation} 
be the matrix whose columns are the vectors $\psi_y$. 
Denote as $\mathcal{A}$, $\mathcal{B}$ the subspaces of $\mathcal{H}$ spanned by the columns of $A$ and by the columns of $B$, respectively, and  
define the reflection operators on $\mathcal{H}$ with respect to $\mathcal{A}$ and $\mathcal{B}$:
\begin{equation} 
{\mathcal R}_A=2AA^H-I,\qquad {\mathcal R}_B=2BB^H-I,\label{eq:reflectors}
\end{equation}
where $I$ is the identity matrix of size $n^2$.
Our quantum walk operator on $\mathcal{H}$ is then defined as
\begin{equation}
W_{P}={\mathcal R}_B{\mathcal R}_A.\label{eq:walkop}
\end{equation}
Applying $W_P$ to a state in $\mathcal{H}$ amounts to performing a step of the quantum walk starting from that state.
 
In the following we will denote as $\Pi_A=AA^H$, $\Pi_B=BB^H$ the projectors with respect to $\mathcal{A}$ and $\mathcal{B}$, respectively.

\begin{definition}
With the setup outlined above, the associated {\em discriminant matrix} is $D=A^HB$, which can be equivalently defined as $D=(D_{xy})$ with $D_{xy}=\sqrt{P_{xy}P^*_{yx}}$. 
\end{definition}
Note that, for a reversible walk, the matrix $D$ is symmetric.

\begin{remark}
It holds
$$
D={\rm diag}(\pi)^{1/2} P\, {\rm diag}(\pi)^{-1/2},
$$
therefore $D$ is always similar to $P$. 
\end{remark}

\begin{remark}
    The action of matrices $A$ and $B$ preserves vector 2-norm. In particular, if $v\in\mathbb{C}^n$ is such that $\|v\|_2=1$, then $\|Av\|_2=\|Bv\|_2=1$.
\end{remark}
\begin{remark}\label{rem:singvalD}
    The singular values of $D$ belong to $[0,1]$. Indeed (see, e.g., \cite{PortugalBook}, section 11.4) from the SVD of $D$ we have
    $$
    D=U\Sigma V^H,\qquad D^H=V\Sigma U^H
    $$
    and therefore
    $$
    Dv_i=\sigma_i u_i,\qquad D^H u_i=\sigma_i v_i,\qquad i=1,\ldots,n,
    $$
    where $\sigma_i$ are the singular values of $D$, whereas $u_i,v_i$ denote the left and right singular vectors, for $i=1,\ldots,n$.
    
    By definition it holds $D=A^H B$. If we multiply the above relations from the left by $A$ and $B$, respectively, we obtain
    $$
    \Pi_A B v_i=\sigma_i A u_i, \qquad \Pi_B A u_i=\sigma_i B v_i,\qquad i=1,\ldots,n.
    $$
     Taking 2-norms on both sides of each equation and recalling the norm-preserving property of $A$ and $B$, we obtain the thesis.
\end{remark}

Let us now recall the spectral theorem (see \cite{Szegedy2004} and \cite{magniez2011search}), which describes the eigenstructure of $W_P$ and in particular yields a relation between the singular values of $D$ and the eigenvalues of $W_P$.

\begin{theorem}[Spectral Theorem]\label{thm:spectral}
With the notation introduced above, let $\theta_1,\ldots,\theta_{\ell}\in(0,\frac{\pi}{2})$ be (possibly repeated) angles such that the singular values of $D$ belonging to the open interval $(0,1)$ can be written as $\cos(\theta_k)$, $k=1,\ldots,\ell$. Let $u_k,v_k$, for $k=1,\ldots,\ell$, be the associated left and right singular vectors. 

Then the eigenvalues of $W_P$ with nonzero imaginary part are
$$
e^{-2i\theta_1}, e^{2i\theta_1},\ldots,e^{-2i\theta_\ell},e^{2i\theta_\ell}
$$
and the (non-normalized) associated eigenvectors are
$$
Au_1-e^{-i\theta_1}Bv_1,\, Au_1-e^{i\theta_1}Bv_1,\ldots,
Au_\ell-e^{-i\theta_\ell}Bv_\ell,\, Au_\ell-e^{i\theta_\ell}Bv_\ell.
$$
Moreover:
\begin{enumerate}
    \item on $\mathcal{A}+\mathcal{B}$ the eigenvalues of $W_P$ with nonzero imaginary part are ${\rm e}^{\pm 2i\theta_k}$, for $k=1,\ldots,\ell$,
    \item on $\mathcal{A}\cap\mathcal{B}$, the walk operator $W_P$ acts as the identity and the subspace $\mathcal{A}\cap\mathcal{B}$ is spanned by the left (or right) singular vectors of $D$ with singular value $1$, 
    \item on $\mathcal{A}\cap\mathcal{B}^{\perp}$ and $\mathcal{A}^{\perp}\cap\mathcal{B}$, the walk operator $W_P$ acts as minus the identity; the subspace $\mathcal{A}\cap\mathcal{B}^{\perp}$ is spanned by the left singular vectors of $D$ with singular value $0$ and the subspace $\mathcal{A}^{\perp}\cap\mathcal{B}$ is spanned by the right singular vectors of $D$ with singular value $0$,
    \item $W_P$ has no other eigenvalues on $\mathcal{A}\cap\mathcal{B}$; on $\mathcal{A}^{\perp}\cap\mathcal{B}^{\perp}$ it acts as the identity.
\end{enumerate}
\end{theorem}

\begin{remark}\label{rem:spectral}
    If $(u, v)$ is a singular vector pair of $D$ associated with singular value $\cos(\theta)$, then the subspace spanned by $Au$ and $Bv$ is invariant w.r.t.~the action of $W$. Also note that $\theta$ is the angle between $Au$ and $Bv$.
\end{remark}

\section{Quantum hitting time}\label{sec:quantumht}

\subsection{Quantum hitting time for symmetric walks} 

The notion of quantum hitting time proposed by Szegedy (\cite{Szegedy2004}, see also \cite{PortugalBook}) applies to the case where $P=P^T$ and is formulated as follows.  

As in the classical case, let $\tilde{P}$ be the modified transition matrix, and consider the modified quantum walk operator $W_{\tilde{P}}$ associated with $\tilde{P}$. 
 Recall that, in the hypothesis of symmetric $P$, the associated classical stationary distribution is uniform, i.e., it is a vector of length $n$ with entries all equal to $1/n$. Define the initial state in uniform superposition
\begin{equation}
|\psi_0\rangle=\frac{1}{\sqrt{n}}\sum_{x\in X,y\in Y}\sqrt{P_{xy}}\ket{x,y}=A\ket{\frac{\bf{1}}{\sqrt{n}}}
\in\mathcal{H}.\label{eq:psi0}
\end{equation}

Observe that $|\psi_0\rangle$ is an eigenvector of $W_P$ with eigenvalue $1$, but in general it is not an eigenvector of $W_{\tilde{P}}$. If we apply both the original and the modified quantum walk with initial state $|\psi_0\rangle$, we can expect that the former will be stationary at $|\psi_0\rangle$, whereas the latter will evolve in a nontrivial way. Therefore the distance $\|W_{\tilde{P}}^t |\psi_0\rangle - W_{P}^t |\psi_0\rangle\|$, evaluated w.r.t.~the $2$-norm on $\mathcal{H}$, gives a measure of how far the modified walk has strayed from the original one at step $t$. This motivates the following definition.

\begin{definition}\label{def:QHTsymm}
The quantum hitting time for a symmetric walk is the smallest value of $T$ such that the quantity
    \begin{equation}\label{ft} F(T)=\frac{1}{T+1} \sum_{t=0}^T \|W_{\tilde{P}}^t |\psi_0\rangle - |\psi_0\rangle\|^2,
    \end{equation}
    is $\geq\frac{n-m}{n},$
    where $m=|M|$.
\end{definition}

Several formulations of quantum hitting time are found in the literature, both for Szegedy and for coined quantum walks; see e.g. \cite{kempe2003proceedings, krovi2006hitting, magniez2012hitting}. Note that quantum hitting time may exhibit significantly different behaviour w.r.t.~classical hitting time: for instance, in some cases it can be infinite \cite{krovi2006quantum}. 

\subsection{Quantum hitting time for time-reversible walks}  
Magniez and co-authors \cite{magniez2011search} analyze the notion of quantum hitting time in the more general case of time-reversible walks. Let  ${\bf \pi}$ be the stationary distribution vector, normalized w.r.t. the 1-norm. The initial state is defined as
\begin{equation}
    \ket{\psi_0}=\sum_{x\in X}\sqrt{\pi_x}\ket{\phi_x}=A\ket{\sqrt{\bf \pi}}\label{eq:psi0rev}
\end{equation}
or, equivalently, as
$$
\ket{\psi_0}=\sum_{y\in Y}\sqrt{\pi_y}\ket{\psi_y}=B\ket{\sqrt{\bf \pi}}.
$$

We can again define the function $F(T)$ as in \eqref{ft}, with the new choice of initial state. By analogy with Definition \ref{def:QHTsymm}, we define the quantum hitting time as follows. 
\begin{definition}\label{def:qht}
The quantum hitting time ($QH$) for a time-reversible walk is
$$
QH_{\pi,M}=\min\left\{T:F(T)\geq 1-\sum_{k\in M}\pi_k\right\},
$$
where the initial state for $F(T)$ is chosen as \eqref{eq:psi0rev}.
\end{definition}
Note that this definition is similar to the one for symmetric walks, except that the threshold value for $F(T)$ is expressed w.r.t. to the stationary distribution, which in general is no longer uniform.

\subsection{Modified walk operator}

Let us look in more detail at the relation between the 
 walk operators $W_P$ and $W_{\tilde{P}}$. Recall that $\tilde{P}$ differs from $P$ only for the target rows in the set $M$, and that the walk operator $W_P$ is defined in \eqref{eq:reflectors},\eqref{eq:walkop} as the product of two reflectors.

By substitution we see that, under the hypothesis that $P_{ii}=0, \forall i=1, \ldots, n,$ 
$$
W_{\tilde{P}}={\mathcal R}_B{\mathcal R}_A E,
$$
where $E$ is a suitable matrix which in particular cases turns out to be a reflection.

Denoting by $\tilde{A}$ and $\tilde{B}$ the analogues of \eqref{eq:Adef} and \eqref{eq:Bdef} for the modified walk, we can rewrite these matrices as rank-$m$ corrections of $A$ and $B$ respectively, i.e.
\begin{align}\nonumber
\tilde{A}=A - A\sum_{x\in M} \ket{x} \bra{x} +\sum_{x\in M} \ket{x} \ket{x}\bra{x} \\ \nonumber
\tilde{B}=B - B\sum_{x\in M} \ket{x} \bra{x} +\sum_{x\in M} \ket{x} \ket{x}\bra{x},
\end{align}
Denoting by $\ket{a_{x}}=A\ket{x}$ and $\ket{b_{x}}=B\ket{x}$, and by $A_M=\sum_{x\in M} \ket{a_{x}}\bra{x}$, $B_M=\sum_{x\in M} \ket{b_{x}}\bra{x}$ we have
\begin{align}\nonumber
{\mathcal R}_{\tilde{A}}={\mathcal R}_A-2\left(\sum_{x\in M} \left(\ket{a_{x}} \bra{a_{x}}-(\ket{x}\ket{x})(\bra{x}\bra{x})\right)\right)\\ \nonumber
{\mathcal R}_{\hat{B}}={\mathcal R}_B-2\left(\sum_{x\in M} \left(\ket{b_{x}} \bra{b_{x}}-(\ket{x}\ket{x})(\bra{x}\bra{x})\right)\right)\
\end{align}
Hence we get
$$
W_{\tilde{P}}={\mathcal R}_{B}{\mathcal R}_{A}-2\, B_MB_M^H{\mathcal R}_{A}-2\, {\mathcal R}_{B}A_MA_M^H+ 4\, B_MB_M^HA_MA_M^H.
$$
Setting $U=[{\mathcal R}_BA_M,B_M]$ and $V=[A_M, {\mathcal R}_AB_M]$, and noticing that $WV=U$ we get

$$
W_{\tilde{P}}={\mathcal R}_{B}{\mathcal R}_{A}(I-2VV^H)+ 4\, B_MB_M^HA_MA_M^H.
$$
 With a little manipulation we can rewrite the above equality also as
$$
W_{\tilde{P}}={\mathcal R}_{B}{\mathcal R}_{A}\left(I-2\,V\left[\begin{array}{cc}I_{m}& -2C\\
O_m& I_m\end{array}\right]V^H\right),
$$
with $C=B_M^HA_M$. Note that if $m=|M|=1$ then $B_M^HA_M=\braket{b_{x}|a_x}=0$. Matrix $V=[\ket{a_x},{\mathcal R}_A \ket{b_x}]={\mathcal R}_A \,[\ket{a_x}, \ket{b_x}]$ has orthonormal columns, in fact $\braket{a_x|{\mathcal R}_A b_x}=\braket{a_x{\mathcal R}_A^H| b_x}=\braket{a_x|b_x}=0,$ and then $I-2VV^H=-(2VV^H-I)$ is a reflector which uses as mirror ${\rm span}\{ V\}^\perp$.

\section{Quadratic speedup for  reversible walks}\label{sec:speedup}

In this section we present in detail the proof of how quantum hitting time improves quadratically over its classical counterpart, that is,  
quantum hitting time is upper bounded by a quantity of the order of the square root of the classical hitting time. We follow the proof that was outlined in \cite{Szegedy2004} under the assumption $P=P^T$, but extend it to the more general hypotheses used in \cite{magniez2011search}, namely, time reversibility  of $P$ rather than symmetry.

With the definitions of $\ket{\psi_0}$ and $F(T)$ given in \eqref{eq:psi0rev} and in the subsequent discussion, we have
\begin{equation} \label{FT}
F(T)=\frac{1}{T+1} \sum_{t=0}^T \| W_{\tilde{P}}^t \ket{\psi_0}-\ket{\psi_0}\|^2= 2\,\braket{\psi_0|\psi_0}-
\frac{2}{T+1} \sum_{T=0}^T\braket{\psi_0| W_{\tilde{P}}^t\psi_0}.
\end{equation}
We are therefore interested in studying the orbit of a given vector $z$ under the action of a quantum walk operator. The following lemma \cite{Szegedy2004} gives such a characterization for a general quantum walk operator $W$ on a finite graph; notation is as in Section \ref{sec:background}.

\begin{lemma}\label{lemma:BzT}
Let $z=\sum_{k} \nu_k z_k$ such that $\|z_k\|=1$ and $z_k\in {\rm span} \{Au_k, Bv_k\}$, where $u_k$ and $v_k$ are respectively the left and right singular vectors of the discriminant matrix $D$ corresponding to the singular value $\cos(\theta_k)\in(0, 1)$. Let 
$$
B(z, T)=\frac{1}{T+1}\sum_{t=0}^T \braket{z\mid W^t z}.
$$
Then for $T\ge 64\sum_k \frac{\nu_k^2}{\theta_k}$ we have $B(z, T)\le 1/2.$
\end{lemma}
\begin{proof}
Vector $z_k$ belongs to ${\rm span} \{Au_k, Bv_k\}$ which is an invariant subspace for $W$. Note that the action of $W$ on ${\rm span} \{Au_k, Bv_k\}$ is a rotation by an angle $2\theta_k$, and the projection of $W z_k$ in the direction of $z_k$ has length $\cos(2\theta_k)$. 
Then $\braket{z_k\mid W^t z_k}=\cos(2t\theta_k),$ and we get
$$
B(z_k, T)=\frac{1}{T+1} \sum_{t=0}^T \cos(2t\theta_k)=\frac{\cos(2T\theta_k)-\cos(2(T+1)\theta_k)+1-\cos(2\theta_k)}{2(T+1)(1-\cos(2\theta_k))}. 
$$
Using the following trigonometric inequalities 
\begin{eqnarray*}
|\cos(\beta)-\cos(\alpha)|\le |\beta-\alpha|,\\
1-\cos(\alpha)\ge \alpha^2/8.
\end{eqnarray*}
we deduce the bound 
$$\frac{\cos(2T\theta_k)-\cos(2(T+1)\theta_k)+1-\cos(2\theta_k)}{2(T+1)(1-\cos(2\theta_k)}\le \min\{1, \frac{4}{(T+1)\theta_k}\}.$$
Note that the minimum in the right-hand side in the equation above is $\frac{4}{(T+1)\theta_k}$ for $T\geq \frac{4}{\theta_k}$.

\begin{equation}\label{Bzt}
    B(z, T)= \sum_{k} \nu_k^2 B(z_k, T)\le \sum_{k} {\nu_k^2} \min\{1, \frac{4}{(T+1){\theta_k}}\}.
\end{equation}
Let $E= \sum_{k} \nu_k^2/\theta_k$ and let $K=\{ k\mid \frac{1}{\theta_k}>4 E\}$. We have
$E=\sum_{k\in K} \frac{\nu_k^2}{\theta_k}+\sum_{k\not \in K} \frac{\nu_k^2}{\theta_k}$, and
$$
4E \sum_{k\in K}\nu_k^2 < \sum_{k\in K} \frac{\nu_k^2}{\theta_k}=E-\sum_{k\not \in K} \frac{\nu_k^2}{\theta_k}.
$$
We derive that $\sum_{k\in K} \nu_k^2<\frac{1}{4E}(E-\sum_{k\not \in K} \frac{\nu_k^2}{\theta_k})< 1/4.$ We then have

$$
B(z, T)\le \sum_{k\in K} \nu_k^2 \min\{1, \frac{4}{(T+1)\theta_k}\}+\sum_{k\not \in K} \nu_k^2 \min\{1, \frac{4}{(T+1)\theta_k}\}\le 
$$
$$
\le \sum_{k\in K} \nu_k^2+ \sum_{k\not \in K} \nu_k^2 \frac{4}{(T+1)\theta_k} < \frac{1}{4}+\frac{16 E}{T+1}.
$$ 
Choosing $T\geq 64\sum_k \frac{\nu_k^2}{\theta_k}$ and recalling that $T$ is an integer, we get $B(z, T)\le 1/2$ as claimed.
\end{proof}

From now on we will need the hypothesis of time reversibility, which ensures in particular that the discriminant matrix $D$ is symmetric.

We can decompose $\ket{\psi_0}$ in equation~\eqref{eq:psi0rev} as the sum of two orthogonal vectors, one belonging to the subspace spanned by the columns of $A$ with indices not in $M$, and one generated by the columns whose indices are in $M$. Define the (non-normalized) vectors $\ket{\psi_{-M}}$ and $\ket{\psi_{M}}$ as follows:

\begin{eqnarray} \label{psidec}
\ket{\psi_{-M}}&=& \sum_{x\not\in M,y}\sqrt{\pi_x} \sqrt{P_{xy}} \ket{x,y},\\ \nonumber
\ket{\psi_{M}}&=& \sum_{x\in M,y} \sqrt{\pi_x} \sqrt{P_{xy}} \ket{x,y}.
\end{eqnarray}

Clearly it holds
$\ket{\psi_{0}}= \ket{\psi_{-M}}+\ket{\psi_{M}}$.
We can prove the following useful facts:
\begin{enumerate}
    \item $\braket{\psi_{-M}\mid \psi_{M}}=0$.
    \item $\braket{\psi_{M}\mid \psi_{M}}= p$ where $p=\sum_{x\in M} \pi_x.$
    Indeed we have
 \begin{eqnarray*}
    \braket{\psi_{M}\mid \psi_{M}}&=&\left(\sum_{x\in M, y}\sqrt{\pi_x}\sqrt{P_{xy}}\bra{xy}\right)\left(\sum_{r\in M, s} \sqrt{\pi_r}\sqrt{P_{rs}}\ket{rs}\right) =\\
    &=&\sum_{x\in M, y}\pi_x P_{xy}\braket {xy\mid xy}= \sum_{x\in M}\pi_x \sum_y P_{xy}=\sum_{x\in M} \pi_x=p. 
    \end{eqnarray*}
    
    \item $\braket{\psi_{-M}\mid \psi_{-M}}=\braket{\psi_{0}\mid \psi_{0}}-\braket{\psi_{M}\mid \psi_{M}}=1-p$.
\end{enumerate}

The discriminant matrix for the modified walk operator has a block diagonal structure
$$
\tilde{D}=\tilde{A}^H \tilde{B}=\begin{bmatrix}  I_m&O\\O & D_{-M}
\end{bmatrix}, \quad D_{-M}\in \mathbb{R}^{(n-m)\times (n-m)}.
$$
Note that the trailing  principal matrix $D_{-M}$ appears also as the trailing  $(n-m)\times(n-m)$ principal block of $D$. In fact, $D$ and $\tilde D$ differ only in the first $m$ rows and $m$ columns. Note that in the case of reversible walks $\tilde D$ is symmetric and hence diagonalizable by a  unitary transformation.

Let 
$$
D=\begin{bmatrix}
D_{11} & D_{12}\\ D_{12}^T & D_{-M}
\end{bmatrix}, \quad D_{11}\in \mathbb{R}^{m\times m}.
$$
Denote by $\hat I_m=[I_m, O]^T$, the $n\times m$ matrix where $O$ is the null matrix of size $(n-m)\times m$ and set $U=D\, \hat I_m-\frac{1}{2}\hat I_m \,(D_{11}+I_m)$. We have $\tilde{D}=D-(\hat I_m U^T+U \hat I_m^T)$.

Now, take the SVD of  $D=V\Gamma V^T$, where $\Gamma={\rm diag}(\gamma_1, \gamma_2, \ldots, \gamma_n)$ with $1=\gamma_1\ge \gamma_2 \ge \cdots \ge \gamma_n\ge 0$. 
We have 
$$\tilde{D}=V(\Gamma- V^T(\hat I_m U^T+U \hat I_m^T)V  )V^T.$$ 
Consider the SVD of $\Gamma- V^T(\hat I_m U^T+U \hat I_m^T)V =Q\Lambda Q^T$. We get
$$
\tilde D= V(\Gamma-V^T(\hat I_m U^T+U \hat I_m^T)V)V^T=V(Q\Lambda Q^T)V^T=S\Lambda S^T,
$$
where the columns of  $S=VQ$ are the singular vectors of $\tilde D$. Let $\{q_k\}_{k=1}^n$, $\{v_k\}_{k=1}^n$ and $\{s_k\}_{k=1}^n$ be the columns of $Q$, $V$ and $S$, respectively. Because of the block diagonal structure of $\tilde D$, we have that $s_i=e_i$, for $i=1, \ldots, m$, while the remaining singular vectors are such that $\hat I_m^T s_k=0$, for $k=m+1, \ldots, n$.
Denoting by $1=\lambda_1=\cdots=\lambda_m> \lambda_{m+1}\ge \cdots \ge \lambda_n$ the eigenvalues of $\tilde D$, we have that the  diagonal entries  of $\Lambda$ are $|\lambda_i|$, $i=1,\ldots,n$.  We have that 
$$(\Gamma- V^T(\hat I_m U^T+U \hat I_m^T)V)q_{m+1}=|\lambda_{m+1}|\,q_{m+1},$$ 
and then 
$$(\Gamma-|\lambda_{m+1}| I)q_{m+1}= V^T\hat I_m U^T V q_{m+1},$$ 
since $\hat I_m V q_{m+1}=\hat I_m^T s_{m+1}=0$.

Let $\Phi= \tilde{A} S $ and $\Psi=\tilde{B} S$. Set $\nu=S^T [0,\sqrt{\pi_{-M}}]^T$: in other words, we are defining $\nu$ such that $\sqrt{\pi_{-M}}$ can be written as a linear combination of the singular vectors of $\tilde{D}$ with coefficients given by the entries of $\nu$. Recall from \eqref{psidec} the expression of $\ket{\psi_{-M}}$ as 
$$
\ket{\psi_{-M}}=\tilde{A}\left[\begin{array}{c}
 0   \\
 \sqrt{\pi_{-M}}\end{array}\right],$$
where the zero block in the right-hand side has length $m$. It holds 
$$\ket{\psi_{-M}}=\tilde A S \nu=\sum_{k=m+1}^n \nu_k \Phi_k$$
or equivalently
$$\ket{\psi_{-M}}=\tilde B S \nu=\sum_{k=m+1}^n \nu_k \Psi_k.$$
 Note that, for $k=m+1,\ldots,n$,
 the subspace $\mathcal{S}_k={\rm span}\{\Phi_k, \Psi_k\}$ is invariant under $W_{\tilde P}$, i.e $W_{\tilde P} \Phi_k \in \mathcal{S}_k$ and $W_{\tilde P} \Psi_k \in \mathcal{S}_k$. A more general proof of this fact will be spelled out in detail in Section \ref{sec:generalized}.
Also note in passing the pseudo-orthogonality relation  
 $\Phi^*\Psi=\Lambda.$

We are now ready to prove a theorem that bounds the quantum hitting time in terms of the eigenvalues of $P$.

\begin{lemma}\label{lemma:FT-inequality}
Let $p=\sum_{x  \in M} \pi_x$, then 
choosing $T\ge \frac{64}{1-p} \sum_{k\not \in M} \frac{\nu_k^2 }{\theta_k}$ we have that $F(T)\ge 1-p.$ 
\end{lemma}
\begin{proof}
From~\eqref{FT} we have
$$
F(T)=\frac{1}{T+1}\sum_{t=0}^T \|W_{\tilde P}^t \ket{\psi_0}-\ket{\psi_0}\|^2=2-2 B(\psi_0, T).
$$
Moreover, it holds
\begin{equation}
(T+1)\,B(\psi_0, T)=\sum_{t=0}^T \left(\braket{\psi_{-M}\mid W_{\tilde P}^t \psi_{-M}} +\braket{\psi_{M}\mid W_{\tilde P}^t \psi_{M}}\right),\label{eq:4terms}
\end{equation}
since for every $t\ge 0$ we have $\braket{\psi_{M}\mid W_{\tilde P}^t \psi_{-M} }+\braket{\psi_{-M}\mid W_{\tilde P}^t \psi_{M} }=0$. The discussion above implies that $\ket{\psi_{-M}}=\sum_{k \not\in M} \nu_k \Phi_k$ belongs to a combination of invariant subspaces under $W_{\tilde P}$. Hence  applying Lemma~\ref{lemma:BzT} to $z=\frac{\psi_{-M}}{\sqrt{1-p}}$, we have 
$$\frac{1}{T+1}\sum_{t=0}^T \braket{\psi_{-M}\mid W_{\tilde P}^t \psi_{-M} }\le \frac{1-p}{2}$$ 
for $T\geq \frac{64}{1-p}\sum_{k \not\in M} \frac{\nu_k^2}{\theta_k}$.

Let us bound the last term in the right-hand side of \eqref{eq:4terms}. The Cauchy-Schwartz inequality yields 
$$
\braket{\psi_{M}\mid W_{\tilde P}^t \psi_{M}}\le \|\psi_{M}\| \,\|W_{\tilde P}^t \psi_{M}]\| =\|\psi_{M}\|^2=p.  
$$

 From \eqref{eq:4terms} we deduce $$
F(T)>1-p
$$
for $T\geq\frac{64}{1-p}\sum_{k \not\in M} \frac{\nu_k^2}{\theta_k}$. 
\end{proof}

The hypothesis of time reversibility ensures that $D$ is symmetric and therefore we have 
$\cos({\theta_k})=|\lambda_k|$. We can then prove the main result about quadratic quantum speedup~\cite{Szegedy2004,magniez2011search}.

\begin{theorem}\label{thm:quadspeedup}
  For time-reversible walks,  it holds
    $$QH_{\pi,M}\leq\frac{64}{\sqrt{1-p}}\sqrt{H_{\pi,M}},$$
    where $p=\sum_{k\in M}\pi_k$ and $\pi$ is the stationary distribution.
\end{theorem}
\begin{proof}
Recall from Definition \ref{def:qht} that the quantum hitting time is the minimum value of $T$ such that $F(T)\geq 1-p$. Therefore, from Lemma \ref{lemma:FT-inequality} we have that the quantum hitting time must be bounded from above by $\frac{64}{1-p} \sum_{k\not \in M}  \frac{\nu_k^2}{\theta_k}$. 
Since it holds in general $1-\cos(\theta_k)\leq \theta_k^2$, we have the bound
    $$
    \frac{1}{\theta_k}\leq\frac{1}{\sqrt{1-|\lambda_k|}}
    $$ 
    and therefore
    $$
    QH_{\pi,M}\leq \frac{64}{1-p} \sum_{k\not \in M}  \frac{\nu_k^2}{\theta_k}\leq 
    \frac{64}{1-p} \sum_{k\not \in M}  \frac{\nu_k^2}{\sqrt{1-|\lambda_k|}}.
    $$
    
    Using, e.g., Jensen's inequality and the fact that $\sum_{k\not\in M}\nu_k^2=1-p$, we see that
    $$
    \sum_{k\not\in M}\frac{\nu_k^2}{\sqrt{1-|\lambda_k|}}\leq\sqrt{1-p}\sqrt{\sum_{k\not\in M}\frac{\nu_k^2}{1-|\lambda_k|}}
    $$
    We can now bound the quantum hitting time as follows:
$$
QH_{\pi,M}\leq \frac{64}{\sqrt{1-p}}\sqrt{\sum_{k\not\in M}\frac{\nu_k^2}{1-|\lambda_k|}}=
\frac{64}{\sqrt{1-p}}\sqrt{H_{\pi, M}},
 $$   
 where we have used \eqref{eq:Hsigma} for the expression of the classical hitting time,  the fact that $P_{-M}= {\rm diag}(\pi_{-M})^{-1/2}D_{-M} {\rm diag}(\pi_{-M})^{1/2}$, and that $\nu=S^T[0, \sqrt{\pi_{-M}}]^T.$ 
 
\end{proof}

Theorem \ref{thm:quadspeedup} implies $QH_{\pi,M}=O(\sqrt{H_{\pi,M}})$. As we will see in Section \ref{sec:experiments}, in practice the bound given by Theorem \ref{thm:quadspeedup} is often pessimistic, since in numerical tests it typically holds $QH_{\pi,M}<\sqrt{H_{\pi,M}}$.

\section{Quantum hitting time for a general distribution}\label{sec:generalized}
The construction in \cite{magniez2011search} relies on the transition matrix $P^*$ associated with the reversed walk:
$$
P_{xy}\pi_x=\pi_y P^*_{y, x},
$$
or equivalently $P^*={\rm{diag}}(\pi)^{-1} P^T {\rm{diag}(\pi)}$,
where $\pi=(\pi_x)_{x\in X}$ is the unique stationary distribution of the ergodic Markov chain defined by $P$. 

Here we investigate alternative definitions of quantum hitting time where the stationary distribution is replaced by a more general one. The purpose is to define a quantum hitting time according to a given distribution $\sigma$ analogously to the definition of the classic hitting time in~\eqref{eq: HTsigma}. 
 Let $\sigma=[\sigma_1, \ldots,\sigma_n]^T$ be a vector with $\sigma_i>0$ for $i=1,\ldots,n$ and $\sum_{i=1}^n\sigma_i=1$. Let us introduce the matrix 
$$
\hat P={\rm diag}(\sigma)^{-1} P^T {\rm diag}(\sigma),
$$
which coincides with $P^*$ when $\sigma=\pi$.

In the general case where $\sigma\neq\pi$, matrix $\hat P$ is not row-stochastic. Let us normalize its rows and define
$$
\mathcal{P}^*=D_r^{-1}\hat P,
$$
where $D_r={\rm diag}(\sum_j\hat{P}_{1,j},\ldots,\sum_j\hat{P}_{n,j})$. Note that $\mathcal{P}^*=P^*$ if and only if $\sigma=\pi$, the stationary distribution for $P$.

We assume here that the graph has no dangling nodes, so that all nodes have positive outdegree. The matrix $\mathcal{P}^*$ describes a ``generalized reversed walk''. 

As above, define the discriminant matrix $\mathcal{D}$ as 
$$
\mathcal{D}_{xy}=\sqrt{P_{xy}\mathcal{P}^*_{yx}}.
$$
Note that $\mathcal{D}$ is no longer similar to $P$. However $\hat{D}$ is similar to $PD_{r}^{-\frac{1}{2}}$, it holds
\begin{equation} \label{dsigma}
\mathcal{D}={\rm diag}(\sigma)^{\frac12} P D_{r}^{-\frac12} {\rm diag}(\sigma)^{-\frac12}.
\end{equation}

In this new setting, the operator $A$ does not change, whereas the operator $B$ needs to be redefined consistently with $\mathcal{P}^*$. Consider the vectors
$$
\ket{\xi_y}=\ket{\mathcal{p}^*_{y}}\otimes\ket{y}=\sum_{x\in X}\sqrt{\mathcal{P}^*_{yx}}\ket{x}\ket{y},
$$
and define $\hat{B}=(\xi_y)_{y\in Y}$ as the matrix whose columns are the vectors $\xi_y$. 

Denote as $\mathcal{\hat{B}}$ the subspace of $\mathcal{H}$ spanned by the vectors $\xi_y$. It holds
$$
\mathcal{D}=A^H\hat{B}.
$$
Note that the argument in Remark \ref{rem:singvalD} also applies to $\mathcal{D}$, with $\hat{B}$ in place of $B$, therefore it is still true that the singular values of $\mathcal{D}$ belong to $[0,1]$. 

The new walk operator is $\hat{W}=\mathcal{R}_{\hat{B}}\mathcal{R}_A=(2\hat{B}\hat{B}^H-I)(2AA^H-I).$
Theorem \ref{thm:spectral} again holds in this setting and ensures the existence of a stationary vector in $\mathcal{A}\cap\mathcal{\hat{B}}$, which can be written explicitly as
$$
\ket{\xi_0}=\sum_{x, y}\sqrt{\sigma_x}\sqrt{P_{xy}}\ket{x}\ket{y}=A\,\ket{\sqrt
{\sigma}}
$$
or equivalently as
$$
\ket{\xi_0}=\hat B D_r^{1/2}\ket{\sqrt{\sigma}}.
$$
Note that, for $\sigma=\pi$, this definition agrees with the one given in~\eqref{eq:psi0rev} for reversible walks. 
Analogously, we can define 
$$
F_{\sigma}(T)=\frac{1}{T+1} \sum_{t=0}^T \|\hat{W}_{\hat P}^t\ket{\xi_0}-\ket{\xi_0}\|_2,
$$
and the Quantum Hitting time according to $\sigma$ as
\begin{equation} \label{eq:qhtsigma}
QH_{\sigma, M}=\min\{T: F_{\sigma}(T)>1-\sum_{k\in M} \sigma_k\}.
\end{equation}

Lemma \ref{lemma:BzT} can be rewritten for this case without major changes.

We would now like to adapt the proof of Lemma \ref{lemma:FT-inequality}.
Consider the discriminant matrix $\mathcal{D}$
and let $\tilde D$ be the discriminant matrix associated with the corresponding absorbing walk, i.e.
$$\tilde{\mathcal{ D}}=\begin{bmatrix}
I_m & 0\\
0& \mathcal{D}_{-M}
\end{bmatrix}.$$
Let
$$
\tilde{\mathcal{ D}}=S\Xi T^H
$$
be an SVD of $\tilde{\mathcal{D}}$. Let us define 
$$
\Phi=AS,\qquad \Psi=\hat{B}T.
$$
Splitting the vector $\ket{\xi_0}$ as the sum  $\ket{\xi_0}=\ket{\xi_{M}}+\ket{\xi_{-M}}$, where $\ket{\xi_{M}}$ is the component of $\ket{\xi_0}$ along the marked states, i.e., 
$$\ket{\xi_M} = \sum_{j}\sum_{k\in M}\sqrt{\sigma_k}\sqrt{P_{kj}}\ket{k}\ket{j}=A\begin{bmatrix}
    \sqrt{\sigma_M} \\ 0
\end{bmatrix}
,$$ 
and $\ket{\xi_{-M}}$ is the component of $\xi_0$ along the unmarked states, i.e., 
$$\ket{\xi_{-M}} = \sum_{j}\sum_{i\not \in M}\sqrt{\sigma_i}\sqrt{P_{ij}}\ket{i}\ket{j}=A\begin{bmatrix}
    0\\\sqrt{\sigma_{-M}}
\end{bmatrix}.$$
The following properties hold:
\begin{enumerate}
    \item $\braket{\xi_M|\xi_{-M}}=0$,
    \item $\braket{\xi_M|\xi_M}=\sum_{k\in M}\sigma_k$,
    \item $\braket{\xi_{-M}|\xi_{-M}}=\sum_{j\not \in M}\sigma_j=1-\sum_{k\in M}\sigma_k$.
\end{enumerate}

 Define vector $\nu$ as 
 \begin{equation}
     \nu=S^T[0; \sqrt{\sigma{-M}}]^T.\label{eq:newnu}
 \end{equation}  
 It holds 
$$
\ket{\xi_{-M}}=\sum_{k\not \in M}\nu_k\Phi(:,k).
$$ 
as a consequence of $[0;\sqrt{\sigma_{-M}}]^T=S\nu.$


Each subspace $\mathcal{S}_k:={\rm span} (\Phi(:,k),\Psi(:,k))$ is invariant under the action of $\hat{W}_P$. Indeed, let us apply $\hat{W}_P$ to $\Phi(:,k)$:
\begin{eqnarray*}
&& W_P \Phi(:,k) = 
(2\hat{B}\hat{B}^H-I) (2 {A}{A}^H-I){A} S(:,k)=\\
&& (2\hat{B}\hat{B}^H-I) (2 {A}-{A}) S(:,k)=\\
&& = (2\hat{B}\hat{B}^H{A}-{A})S(:,k)=\\
&& = (2\hat{B} D_1^H -{A} ) S(:,k) = \\
&& = (2\hat{B} T\Xi S^H -{A} ) S(:,k)=\\
&& = (2 \Psi \Xi S^H -A) S(:,k)=\\
&& = 2\Psi \Xi e_k - \Phi(:,k)=\\
&& = 2\Psi \cos(\theta_k) e_k - \Phi(:,k)=\\
&& = 2\cos(\theta_k)\Psi(:,k) - \Phi(:,k),
\end{eqnarray*}
which belongs to $\mathcal{S}_k$.

A similar computation shows $\hat{W}_P \Psi(:,k)\in\mathcal{S}_k$.

We can now apply Lemma \ref{lemma:BzT} to $\ket{\xi_{-M}}$ and conclude that $B(\xi_{-M},T)<\frac12$ for a sufficiently large $T$. 

Following a similar argument as in Lemma \ref{lemma:FT-inequality} we obtain an upper bound for the generalized quantum hitting time.

\begin{theorem} \label{thm:FT:inequality}
    With the notation introduced above, it holds
    $$
    QH_{\sigma, M}\leq\frac{64}{1-p}\sum_{k\not\in M}\frac{\nu_k^2}{\tilde{\theta}_k},
    $$
    where the coefficients $\nu_k$ are defined as in \eqref{eq:newnu} and the angles $\tilde{\theta}_k\in(0,\pi/2)$ are such that $\cos(\tilde{\theta}_k)$, for $k=1,\ldots,n-m$ are the singular values of ${\mathcal{D}}_{-M}.$
\end{theorem}

We have already seen in Section \ref{sec:background} that the classical hitting time can be defined in terms of a general distribution.  For a distribution $\sigma$ and a set $M$ of marked vertices it holds
$$
H_{\sigma, M}= \sigma_{-M }^T(I-P_{-M})^{-1} {\bf 1}_{- M}.
$$
Substituting to $P_{- M}$ the analogous expression for ${\mathcal{D}}$ in~\eqref{dsigma} we have
\begin{eqnarray*}
H_{\sigma, M}&=& \sigma_{- M}^T {\rm diag}({\sigma_{-M}})^{-1/2}(I-\mathcal{D}_{- M}\hat{D}_r^{1/2})^{-1} {\rm diag}({\sigma_{-M}})^{1/2}{\bf 1}_{- M}=\\
&=&\sqrt{\sigma_{-M}}^T (I-\mathcal{D}_{- M}\hat D_r^{1/2})^{-1}\sqrt{\sigma_{-M}},
\end{eqnarray*}
where $\hat{D}_r$ denotes the principal submatrix obtained by $D_r$ taking only non-marked indices.

Note that $(I-{\mathcal{D}}_{- M}\hat{D}_r^{1/2})$ is an M-matrix implying $(I-{\mathcal{D}}_{- M}\hat{D}_r^{1/2})^{-1}\ge 0.$ Hence the hitting time is always nonnegative.

Also note that $P_{-M}$ and $\mathcal{D}_{- M}{\hat D}_r^{1/2}$ are similar, hence they have the same eigenvalues denoted by  $1>|\lambda_{m+1}|\ge|\lambda_{m+2}|\ge \cdots\ge |\lambda_n|$. We can rewrite the classical hitting time as follows
\begin{equation} \label{eq: HTsigma}
H_{\sigma,  M}=\sum_{i=1}^{n-m} \frac{x_i y_i}{1-\lambda_{i+m}},
\end{equation}
where $\mathcal{D}_{- M}D_r^{1/2}= V \Lambda V^{-1}$, $\Lambda= {\rm diag}(\lambda_{m+1}, \lambda_{m+2}, \ldots, \lambda_n),$ and we denoted  $x=V^T\sqrt{\sigma_{-M}}$ and $y=V^{-1}\sqrt{\sigma_{- M}}.$

In general we cannot easily bound the right hand side of ~\ref{eq: HTsigma}  because it the sum may contain also negative or complex terms. However, the Perron-Frobenius theorem ensures $x_1 y_1>0$, and we conjecture that, at least for some families of graphs,  we have  
$$
\frac{x_1y_1}{1-\lambda_{m+1}}\le H_{\sigma, M} \le \frac{1}{1-\lambda_{m+1}}\sum_{i=1}^{n-m} x_i y_i= \frac{1-p}{1-\lambda_{m+1}}.
$$
If this is the case, we have quadratic speedup whenever $\theta_{\min}\ge \sqrt{1-\lambda_{m+1}}.$

\section{Numerical experiments}\label{sec:experiments}
The purpose of this section is twofold. First, we want to show that even for time-reversible graphs, such as undirected graphs, the bound given by Theorem~\ref{thm:quadspeedup} is not tight, and in general, it is verified experimentally that $QH_{\pi, i}\le \sqrt{H_{\pi, i}}$. There is considerable difference between the computed quantum hitting time and the upper bound given by Lemma~\ref{lemma:FT-inequality}: indeed the constant multiplying $\sqrt{H_{\pi, i}}$ in Theorem~\ref{thm:quadspeedup} is often large and, for graphs with moderate hitting time, it is predominant. The second goal is to show that, although in the case of a generic distribution or for non-reversible graphs we have not proved quadratic speedup, experimental evidence shows that in general quantum walks have much faster convergence than the analogous classic walks,  except in very special cases. 

Below we report the results obtained on different families of graphs built using NetworkX~\cite{SciPyProceedings_11}, a Python package for the creation, manipulation, and study of the structure, dynamics, and functions of complex networks.

We considered the following families of graphs:
\begin{itemize}
\item Circulant graphs of different sizes where each node has a self loop. 

\item Barab\'asi-Albert graphs, that is, undirected graphs built according to a preferential attachment procedure. The associated NetworkX function is {\tt barabasi\_albert\_graph(n,m)}, where $n$ is the total number of nodes and $m$ is the number of edges added together with each new node. 

\item Barbell graphs, which are formed by two complete graphs connected by a path. The associated NetworkX function is {\tt barbell\_graph(m1,m2)}, where $m_1$ is the size of the left and right bells, and $m_2$ is the length of the connecting path. 

\item  Erd\"os-R\'enyi random directed graphs. The associated NetworkX function is {\tt erdos\_renyi\_graph(n, p)}, where $n$ is the number of nodes and $p$ is the probability of edge creation. 


\item Random regular graphs.  The associated NetworkX function is {\tt random\_regular\_graph(d, n)} and yields a random undirected, $d$-regular graph with $n$ nodes. 
    
\end{itemize}

In addition to the stationary distribution $\pi$, we tested several other distributions, such as:
\begin{itemize}
\item the uniform distribution $\frac{1}{n} {\bf 1}_n$,
\item degree-weighted distributions, i.e., defined according to the normalized in/out-degree,
\item the $\varepsilon$-stationary distribution obtained as an $\varepsilon$-perturbation of the stationary distribution, i.e., $\sigma = |\pi ({\bf 1}_n + \varepsilon\, \mbox{randn}(n))|$ and then normalizing the vector to sum one. In our test we use $\varepsilon=10^{-2}$,
\item a Dirac-like distribution concentrated in the first node; this is built by assigning weight 1 to the first node, weight $\delta$ to all other nodes, and normalizing the distribution vector to unitary $1$-norm. Parameter $\delta$ is chosen here as $10^{-2}$.
\end{itemize}

In our tests, we considered the case where the set of marked vertices consists of only one element, and we  compared quantum and classical hitting time by varying the marked vertex from 1 to $n$, the number of nodes in the considered graph.

We provide here statistical results for our numerical tests. In particular we consider the following measures.
Let us denote by 
$$
MQH_{\sigma}=\frac{1}{n}\sum_{i=1}^n QH_{\sigma, i},
$$ 
the average value of the quantum hitting time obtained by taking,  in turn, each node as a marked vertex. For each node $i$ the quantum hitting time $QH_{\sigma, i}$ is calculated as follows. Let $T_0$ be the smallest integer value such that $F(T_0)>1-\sigma_i$ and consider the segment of line joining  points $(T_0-1, F(T_0-1))$ and $(T_0, F(T_0))$. Then the hitting time is chosen as the value of $T$ at which the interpolating line takes the value $1-\sigma_i$; note that this is a (possibly) non-integer number.
The same quantity for classical hitting time is denoted by $MH_{\sigma}$. 

The upper bound of the quantum hitting time given by Lemma~\ref{lemma:FT-inequality}  and Theorem~\ref{thm:FT:inequality} is denoted as $QHE_{\sigma, i}$, whereas  
$$ MQHE_{\sigma}=\frac{1}{n}\sum_{i=1}^n QHE_{\sigma, i}$$ is
the average of estimate of the quantum hitting time.  
We set $$MCHE_{\sigma}=\frac{1}{n}\sum_{i=1}^n \frac{64}{\sqrt{1-\sigma_{i}}}\sqrt{H_{\sigma, i}}, $$ i.e., the mean value of the upper bound to the quantum hitting time given by Theorem~\ref{thm:quadspeedup}, while $$MSH_{\sigma}=\frac{1}{n}\sum_{i=1}^n \sqrt{H_{\sigma, i}}$$ is the mean value of the square root of the classical hitting time over all the nodes.

\begin{table}[]
    \centering
 \begin{tabular}{|r|r|r|r|r|r|r|} \hline
    $n$&$\sigma$& $MQH_{\sigma}$ & $MQHE_{\sigma}$ & $MCHE_{\sigma}$
    & $MSH_{\sigma}$ & $MH_{\sigma}$\\ \hline
    
    25&stationary & 13.067& 238.767&  391.918  & 6.000 & 36.000\\
    &random & 20.597& 297.424&388.913 & 5.994 & 36.000 \\ \hline
   50&  stationary &26.332&  335.519& 554.256& 8.573 & 73.500 \\ 
    &random & 45.153& 437.405& 554.910 & 8.572 &73.500\\ \hline
   75& stationary &39.593& 409.942&  678.823&10.536 & 111.000 \\
  & random &71.671& 549.751& 680.781& 10.533 & 111.000\\ \hline
100&stationary &52.852& 472.761 &783.837& 12.186 & 148.500\\
&random& 77.715& 569.677& 780.280& 12.181&148.500\\ \hline    
\end{tabular}
    \caption{Results for circulant graphs of increasing size, with self loops.  In this case the stationary vector is uniform, and the average classical hitting time does not depend on the initial distribution. Note however that the classical hitting time for the stationary distribution is the same for all nodes, whereas it changes with the marked node when the random distribution is taken, even if the average value is the same. }
    \label{tab:circ}

\end{table}
In Table~\ref{tab:circ} we report the results obtained on graphs with circulant structure plus self loops. Here we tested only the random and the stationary distribution.  In fact for such graphs the stationary distribution is uniform and also coincides with the distributions weighted according to in- and out-degree. This is one of the few examples where $MQH_{\sigma}> MSH_{\sigma}$. However, the inequality of Lemma~\ref{lemma:FT-inequality} still holds, even if the hypothesis for the application of the lemma are not satisfied. Note that in this example both quantum and classical hitting time grow linearly. Classical hitting time stays roughly the same, independently of which distribution is used, and grows with slope $\approx 3/2$. Quantum hitting time grows with slope $\approx 1/2$ for the stationary distribution and $\approx 3/4$ for a random distribution.

In Table~\ref{tab:BA} we report the results obtained on Barab\'asi-Albert graphs for different graph sizes. The graphs in this family are undirected and hence the walk is time-reversible. We are then under the hypotheses of Theorem~\ref{thm:quadspeedup} for the stationary case, i.e.  $\sigma=\pi$, implying quadratic speedup. We note however that experimentally we get a stronger result since not only $MQHE_{\pi}<MCHE_{\pi}$ but also $MQH_{\pi}<MSH_{\pi}$, and such results hold also for the other distributions. Indeed for our test we observe that for each node $i$ we have $QH_{\sigma,i}<\sqrt{H_{\sigma, i}}$. 
Note that the results reported here are the average of the corresponding values over 10 randomly generated graphs of the family.
{\small
\begin{table}[]
	\centering
	\begin{tabular}{|r|l|r|r|r||r|l|r|r|r|}
		\hline
		 $n, m$ & $\sigma$                 & $MQH_{\sigma}$ & $MSH_{\sigma}$ & $MH_{\sigma}$ &$n, m$ & $\sigma$                 & $MQH_{\sigma}$ & $MSH_{\sigma}$ & $MH_{\sigma}$\\ \hline
		  25 & unif.                 &           3.59 &           5.76 &         35.55  &  50 & unif.                  &           4.88 &           7.81 &         64.94 \\
		5        & deg.                  &           3.65 &           5.75 &         35.23   &10  &deg.                   &           4.92 &           7.81 &         64.54 \\
		        & rand.                   &           3.95 &           5.76 &         35.53 & & rand.                  &           5.27 &           7.81 &         64.93  \\
		        & $\varepsilon$-stat. &           3.68 &           5.77 &         35.74 && $\varepsilon$-stat. &           5.05 &           7.82 &         65.13 \\
		        & stat.               &           3.68 &           5.77 &         35.74 && stat.               &           5.05 &           7.82 &         65.13 \\ \hline
		
	          \hline \hline
		 75 & unif.                  &           5.88 &           9.33 &         90.24 &100 & unif.                  &           6.78 &          10.73 &        119.57  \\
		  15      & deg.                   &           5.87 &           9.32 &         90.05  & 20& deg.                   &           6.75 &          10.73 &        119.39 \\
		        & rand.                   &           6.29 &           9.32 &         90.22 & & rand.                   &           7.24 &          10.73 &        119.58 \\
		        & $\varepsilon$-stat. &           6.06 &           9.33 &         90.38 & & $\varepsilon$-stat. &           7.00 &          10.74 &        119.71 \\
		        & stat.               &           6.06 &           9.33 &         90.38 &
	
		        & stat.               &           7.00 &          10.74 &        119.71 \\ \hline
	\end{tabular}
	\caption{Barab\'asi-Albert-graph: average over 10 tests. We see that the quantum and classical hitting time depends only mildly on the initial distribution and we always have $MQH_{\sigma}<MSH_{\sigma}.$ }
	\label{tab:BA}
\end{table}
}

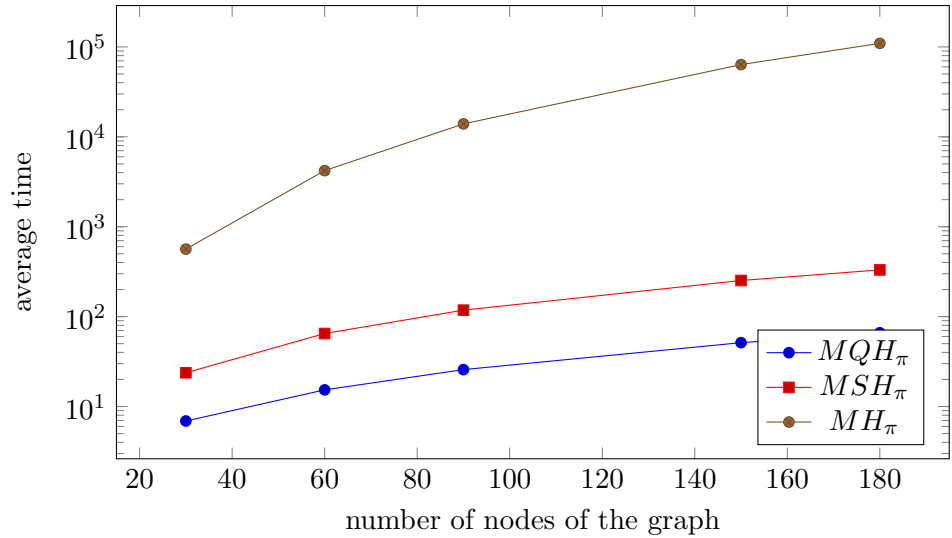
\begin{figure}
\begin{tikzpicture}
\begin{semilogyaxis}[width=\textwidth, height=0.4\textheight, xlabel={number of nodes of the graph}, ylabel={average time}, legend pos=south east]

\addplot table [x=n,y=qht] \loadedtable;
\addplot table [x=n,y=sht] \loadedtable;
\addplot table [x=n,y=ht] \loadedtable;

\legend{$MQH_{\pi}$,$MSH_{\pi}$,$MH_{\pi}$ };
\end{semilogyaxis}
\end{tikzpicture}
\caption{Logarithmic plot  comparing the dependence on the graph size $n$ of mean quantum hitting time, the $MSH$ and mean hitting-time for Barbell graphs of increasing size. } \label{fig:barbell}
\end{figure}

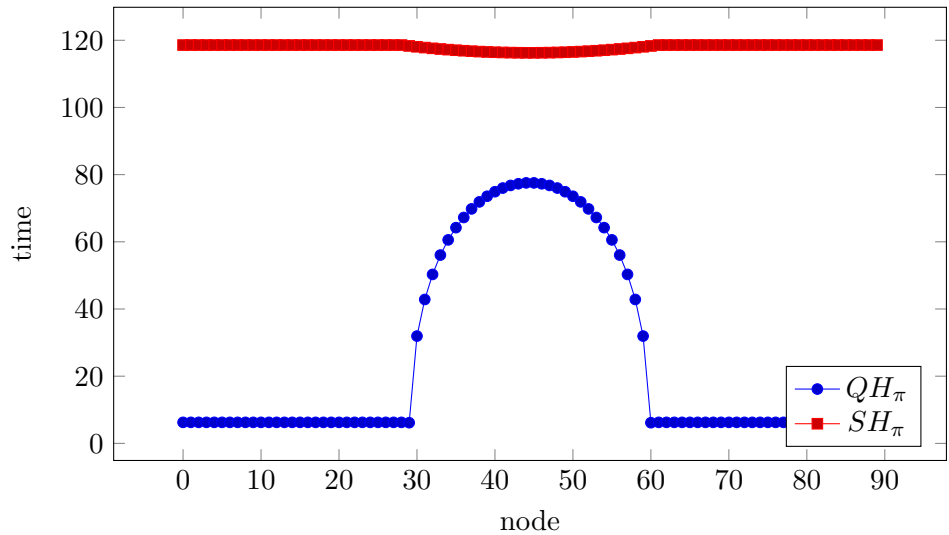
\begin{figure}
\begin{tikzpicture}
\begin{axis}[width=\textwidth, height=0.4\textheight, xlabel={node}, ylabel={time}, legend pos=south east]

\addplot table [x=node,y=qh] \loadedtable;
\addplot table [x=node,y=sht] \loadedtable;

\legend{$QH_{\pi}$,$SH_{\pi}$};
\end{axis}
\end{tikzpicture}
\caption{Plot of quantum and square root of classical hitting times for a barbell graph with $90$ nodes and stationary distribution. It is interesting to note how quantum hitting time distinguishes more clearly among nodes in the ``bar'' (i.e., nodes indexed from $30$ to $60$).} \label{fig:barbell-pi}
\end{figure}

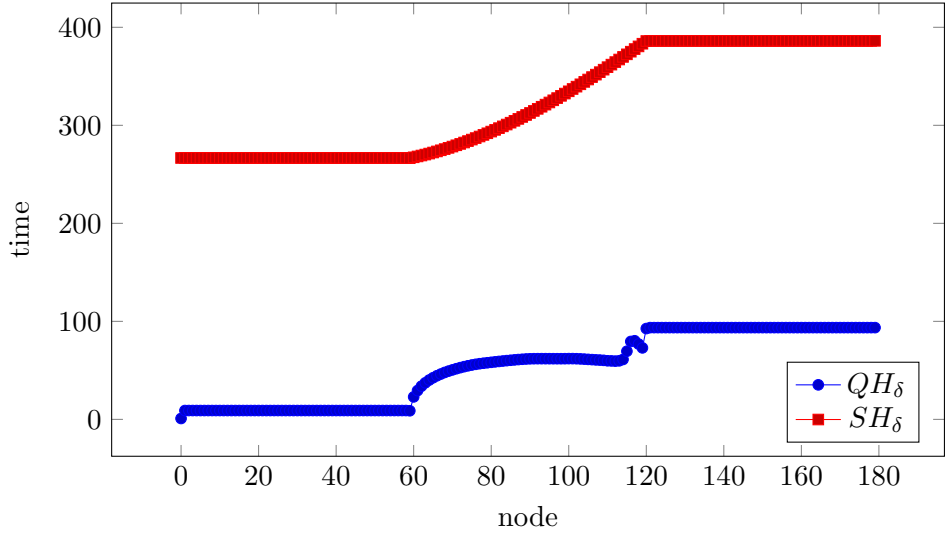
\begin{figure}
\begin{tikzpicture}
\begin{axis}[width=\textwidth, height=0.4\textheight, xlabel={node}, ylabel={time}, legend pos=south east]

\addplot table [x=node,y=qh] \loadedtable;
\addplot table [x=node,y=sht] \loadedtable;

\legend{$QH_{\delta}$,$SH_{\delta}$};
\end{axis}
\end{tikzpicture}
\caption{Plot of quantum and classical hitting times for a barbell graph with $180$ nodes and Dirac-like distribution centered at the first node. Here quantum and square root of classical hitting times display a similar behavior.} \label{fig:barbell-pun}
\end{figure}

The barbell graphs used in these tests have $3n$ nodes: $n$ nodes in each bell, and $n$ nodes in the bar.
Barbell graphs are known \cite{aldous2002reversible} to have a maximal average classical hitting time increasing as $\Omega(n^3)$. It is interesting to compare this behavior with the growth of quantum hitting time. In fact we can observe that quantum hitting time grows only mildly with the size of the graph: see Figure \ref{fig:barbell}, where we plot in a logarithmic scale the average value of the hitting time as the number of nodes increases. Figures \ref{fig:barbell-pi} and \ref{fig:barbell-pun} show quantum and the square root of the classical hitting times for all the nodes of a barbell graph, first w.r.t.~ to the stationary distribution and then w.r.t.~ a Dirac-like distribution. It turns out that quantum hitting time can better differentiate among the nodes in the bar, particularly in the stationary distribution case.

Results for Erd\"os-R\'enyi graphs are reported in Table~\ref{tab:ER}. These are oriented graphs, for which we do not have theoretical results suggesting that quantum walks hit quadratically faster than the corresponding classical walks.    
We note that the classical mean values appear to be essentially independent of the initial walker distribution, while the quantum hitting time is more sensitive to the initial distribution. In particular, when we consider the Dirac-like distribution we get a  higher mean quantum hitting time. This appears to be a typical behavior of quantum walks, which tend to diffuse slowly from a single starting node; see e.g. the discussion in \cite{paparo2014quantum}. Looking at the value of $QH$ for single nodes in the case of the Dirac-like distribution, we can have that $QH_{\delta, i}>\sqrt{H_{\delta, i}}$.

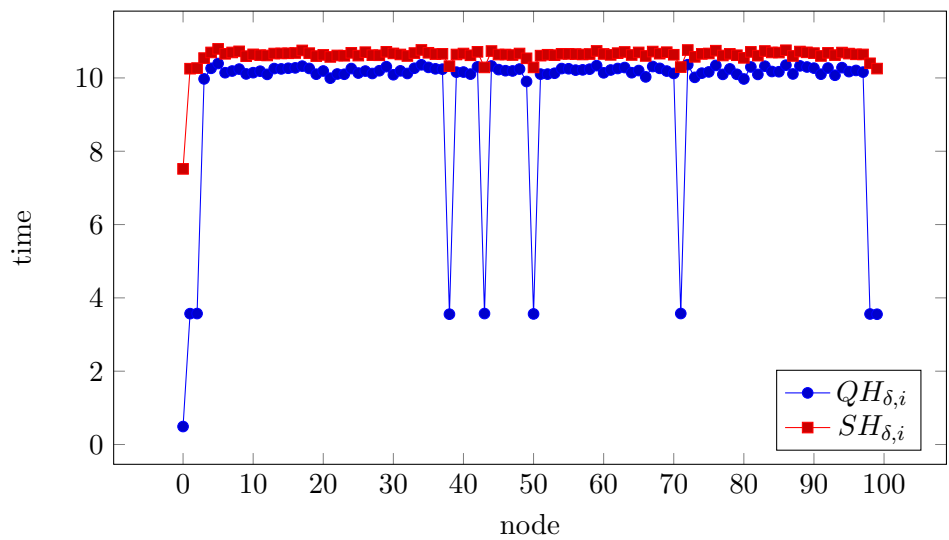
\begin{figure}
\begin{tikzpicture}
\begin{axis}[width=\textwidth, height=0.4\textheight, xlabel={node}, ylabel={time}, legend pos=south east]

\addplot table [x=node,y=qh] \loadedtable;
\addplot table [x=node,y=sht] \loadedtable;

\legend{$QH_{\delta, i}$,$SH_{\delta, i}$};

\end{axis}
\end{tikzpicture}
\caption{Plot of quantum hitting time and square root of classical hitting time for a random-8-regular graph with $100$ nodes and Dirac-like distribution centered at the first node (labeled as node $0$). Note that the 8 nodes adjacent to node $0$ have have a remarkably lower quantum hitting time, whereas the difference is much less marked for the square root of the classical hitting time.} \label{fig:regular-pun}
\end{figure}

Table~\ref{tab:RR} reports the results obtained on undirected $d$-regular random graphs. These results are computed as the average of the corresponding values obtained on 10 randomly generated graphs.
A feature that emerges from the tests is the independence of classical hitting time on the initial distribution: the values of $MH_{\sigma}$ for each graph size are very close. Generally speaking, quantum walks appear to be more sensitive to the initial distribution. It is also interesting to look at the behavior of hitting time for each node, with a Dirac-like distribution (Figure~\ref{fig:regular-pun}). In this example the graph has $100$ nodes, labelled from $0$ to $99$, and each node has $8$ neighbors. The Dirac-like distribution is concentrated on node $0$. Quantum hitting time exhibits marked downward spikes in correspondence of nodes that are adjacent to node $1$, as these nodes can be easily reached from node $1$.
Variations of classical hitting time, on the other hand, are less marked.

\begin{table}[]
	\centering
	\begin{tabular}{|r|l|r|r|r|}
		\hline
		  $n, d$ & $\sigma$                 & $MQH_{\sigma}$ & $MSH_{\sigma}$ & $MH_{\sigma}$ \\ \hline
		 20, 0.6 & uniform                  &           2.76 &           4.35 &         19.32 \\
		         & outdegree                &           2.82 &           4.35 &         19.33 \\
		         & indegree                 &           2.75 &           4.35 &         19.23 \\
		         & random                   &           2.93 &           4.35 &         19.31 \\
		         & $\varepsilon$-stationary &           2.82 &           4.36 &         19.39 \\
		         & Dirac-like               &           4.97 &           4.32 &         19.40 \\
		         & stationary               &           2.82 &           4.36 &         19.39 \\ \hline
		 40, 0.6 & uniform                  &           4.01 &           6.24 &         39.23 \\
		         & outdegree                &           4.03 &           6.24 &         39.23 \\
		         & indegree                 &           4.00 &           6.24 &         39.20 \\
		         & random                   &           4.23 &           6.24 &         39.23 \\
		         & $\varepsilon$-stationary &           4.05 &           6.24 &         39.26 \\
		         & Dirac-like               &           5.81 &           6.22 &         39.20 \\
		         & stationary               &           4.05 &           6.24 &         39.26 \\ \hline
		 50, 0.6 & uniform                  &           4.51 &           6.99 &         49.11 \\
		         & outdegree                &           4.53 &           6.99 &         49.12 \\
		         & indegree                 &           4.50 &           6.99 &         49.09 \\
		         & random                   &           4.75 &           6.99 &         49.10 \\
		         & $\varepsilon$-stationary &           4.54 &           6.99 &         49.13 \\
		         & Dirac-like               &           6.21 &           6.97 &         49.05 \\
		         & stationary               &           4.54 &           6.99 &         49.13 \\ \hline
		100, 0.6 & uniform                  &           6.48 &           9.94 &         99.05 \\
		         & outdegree                &           6.49 &           9.94 &         99.05 \\
		         & indegree                 &           6.48 &           9.94 &         99.04 \\
		         & random                   &           6.82 &           9.94 &         99.05 \\
		         & $\varepsilon$-stationary &           6.50 &           9.94 &         99.06 \\
		         & Dirac-like               &           7.84 &           9.93 &         99.06 \\
		         & stationary               &           6.50 &           9.94 &         99.06 \\ \hline
	\end{tabular}
	\caption{Erd\"os-R\'enyi: the results are the average over 10 tests. We note that the classical  mean values appear to be independent of the initial walker distribution.  }
	\label{tab:ER}
\end{table}

\pgfplotstableread{
n qht  ub   mc  sht  ht
30      6.9000e+00   2.2496e+02   1.5539e+03   2.3730e+01   5.6367e+02
60   1.5320e+01   3.5267e+02   4.2008e+03   6.4860e+01   4.2068e+03
90   2.5740e+01   4.6429e+02   7.6144e+03   1.1801e+02   1.3928e+04
150   5.1300e+01   6.6225e+02   1.6220e+04   2.5220e+02   6.3603e+04
180 66.12 753.12 21270.89 330.99  109557.84

}\loadedtable

\pgfplotstableread{
node qh upq ubh ht sht e1 e1 tr
0 6.254 419.096 7650.659 14060.207 118.576 0.017 0.006 False
1 6.254 419.096 7650.659 14060.207 118.576 0.017 0.006 False
2 6.254 419.096 7650.659 14060.207 118.576 0.017 0.006 False
3 6.254 419.096 7650.659 14060.207 118.576 0.017 0.006 False
4 6.254 419.096 7650.659 14060.207 118.576 0.017 0.006 False
5 6.254 419.096 7650.659 14060.207 118.576 0.017 0.006 False
6 6.254 419.096 7650.659 14060.207 118.576 0.017 0.006 False
7 6.254 419.096 7650.659 14060.207 118.576 0.017 0.006 False
8 6.254 419.096 7650.659 14060.207 118.576 0.017 0.006 False
9 6.254 419.096 7650.659 14060.207 118.576 0.017 0.006 False
10 6.254 419.096 7650.659 14060.207 118.576 0.017 0.006 False
11 6.254 419.096 7650.659 14060.207 118.576 0.017 0.006 False
12 6.254 419.096 7650.659 14060.207 118.576 0.017 0.006 False
13 6.254 419.096 7650.659 14060.207 118.576 0.017 0.006 False
14 6.254 419.096 7650.659 14060.207 118.576 0.017 0.006 False
15 6.254 419.096 7650.659 14060.207 118.576 0.017 0.006 False
16 6.254 419.096 7650.659 14060.207 118.576 0.017 0.006 False
17 6.254 419.096 7650.659 14060.207 118.576 0.017 0.006 False
18 6.254 419.096 7650.659 14060.207 118.576 0.017 0.006 False
19 6.254 419.096 7650.659 14060.207 118.576 0.017 0.006 False
20 6.254 419.096 7650.659 14060.207 118.576 0.017 0.006 False
21 6.254 419.096 7650.659 14060.207 118.576 0.017 0.006 False
22 6.254 419.096 7650.659 14060.207 118.576 0.017 0.006 False
23 6.254 419.096 7650.659 14060.207 118.576 0.017 0.006 False
24 6.254 419.096 7650.659 14060.207 118.576 0.017 0.006 False
25 6.254 419.096 7650.659 14060.207 118.576 0.017 0.006 False
26 6.254 419.096 7650.659 14060.207 118.576 0.017 0.006 False
27 6.254 419.096 7650.659 14060.207 118.576 0.017 0.006 False
28 6.254 419.096 7650.659 14060.207 118.576 0.017 0.006 False
29 6.150 418.021 7633.736 13998.074 118.313 0.017 0.006 False
30 31.938 492.303 7617.358 13938.074 118.060 0.017 0.006 False
31 42.834 515.354 7602.040 13882.074 117.822 0.018 0.006 False
32 50.297 529.853 7587.789 13830.074 117.601 0.018 0.006 False
33 56.058 540.339 7574.610 13782.074 117.397 0.018 0.006 False
34 60.575 548.415 7562.509 13738.074 117.210 0.019 0.007 False
35 64.218 554.844 7551.492 13698.074 117.039 0.019 0.007 False
36 67.245 560.058 7541.562 13662.074 116.885 0.019 0.007 False
37 69.799 564.321 7532.725 13630.074 116.748 0.020 0.007 False
38 71.887 567.807 7524.984 13602.074 116.628 0.020 0.007 False
39 73.567 570.637 7518.342 13578.074 116.525 0.021 0.007 False
40 74.913 572.894 7512.803 13558.074 116.439 0.021 0.008 False
41 76.006 574.639 7508.369 13542.074 116.370 0.022 0.008 False
42 76.785 575.915 7505.041 13530.074 116.319 0.022 0.008 False
43 77.279 576.751 7502.822 13522.074 116.284 0.023 0.008 False
44 77.542 577.164 7501.712 13518.074 116.267 0.024 0.008 False
45 77.542 577.164 7501.712 13518.074 116.267 0.024 0.008 False
46 77.279 576.751 7502.822 13522.074 116.284 0.023 0.008 False
47 76.785 575.915 7505.041 13530.074 116.319 0.022 0.008 False
48 76.006 574.639 7508.369 13542.074 116.370 0.022 0.008 False
49 74.913 572.894 7512.803 13558.074 116.439 0.021 0.008 False
50 73.567 570.637 7518.342 13578.074 116.525 0.021 0.007 False
51 71.887 567.807 7524.984 13602.074 116.628 0.020 0.007 False
52 69.799 564.321 7532.725 13630.074 116.748 0.020 0.007 False
53 67.245 560.058 7541.562 13662.074 116.885 0.019 0.007 False
54 64.218 554.844 7551.492 13698.074 117.039 0.019 0.007 False
55 60.575 548.415 7562.509 13738.074 117.210 0.019 0.007 False
56 56.058 540.339 7574.610 13782.074 117.397 0.018 0.006 False
57 50.297 529.853 7587.789 13830.074 117.601 0.018 0.006 False
58 42.834 515.354 7602.040 13882.074 117.822 0.018 0.006 False
59 31.938 492.303 7617.358 13938.074 118.060 0.017 0.006 False
60 6.150 418.021 7633.736 13998.074 118.313 0.017 0.006 False
61 6.254 419.096 7650.659 14060.207 118.576 0.017 0.006 False
62 6.254 419.096 7650.659 14060.207 118.576 0.017 0.006 False
63 6.254 419.096 7650.659 14060.207 118.576 0.017 0.006 False
64 6.254 419.096 7650.659 14060.207 118.576 0.017 0.006 False
65 6.254 419.096 7650.659 14060.207 118.576 0.017 0.006 False
66 6.254 419.096 7650.659 14060.207 118.576 0.017 0.006 False
67 6.254 419.096 7650.659 14060.207 118.576 0.017 0.006 False
68 6.254 419.096 7650.659 14060.207 118.576 0.017 0.006 False
69 6.254 419.096 7650.659 14060.207 118.576 0.017 0.006 False
70 6.254 419.096 7650.659 14060.207 118.576 0.017 0.006 False
71 6.254 419.096 7650.659 14060.207 118.576 0.017 0.006 False
72 6.254 419.096 7650.659 14060.207 118.576 0.017 0.006 False
73 6.254 419.096 7650.659 14060.207 118.576 0.017 0.006 False
74 6.254 419.096 7650.659 14060.207 118.576 0.017 0.006 False
75 6.254 419.096 7650.659 14060.207 118.576 0.017 0.006 False
76 6.254 419.096 7650.659 14060.207 118.576 0.017 0.006 False
77 6.254 419.096 7650.659 14060.207 118.576 0.017 0.006 False
78 6.254 419.096 7650.659 14060.207 118.576 0.017 0.006 False
79 6.254 419.096 7650.659 14060.207 118.576 0.017 0.006 False
80 6.254 419.096 7650.659 14060.207 118.576 0.017 0.006 False
81 6.254 419.096 7650.659 14060.207 118.576 0.017 0.006 False
82 6.254 419.096 7650.659 14060.207 118.576 0.017 0.006 False
83 6.254 419.096 7650.659 14060.207 118.576 0.017 0.006 False
84 6.254 419.096 7650.659 14060.207 118.576 0.017 0.006 False
85 6.254 419.096 7650.659 14060.207 118.576 0.017 0.006 False
86 6.254 419.096 7650.659 14060.207 118.576 0.017 0.006 False
87 6.254 419.096 7650.659 14060.207 118.576 0.017 0.006 False
88 6.254 419.096 7650.659 14060.207 118.576 0.017 0.006 False
89 6.254 419.096 7650.659 14060.207 118.576 0.017 0.006 False

}\loadedtable

\pgfplotstableread{
node qh upq ubh ht sht e1 e1 tr

0 0.886 391.304 17097.189 71109.900 266.664 0.023 0.002 False
1 9.002 334.877 17102.308 71152.492 266.744 0.023 0.002 False
2 9.002 334.877 17102.308 71152.492 266.744 0.023 0.002 False
3 9.002 334.877 17102.308 71152.492 266.744 0.023 0.002 False
4 9.002 334.877 17102.308 71152.492 266.744 0.023 0.002 False
5 9.002 334.877 17102.308 71152.492 266.744 0.023 0.002 False
6 9.002 334.877 17102.308 71152.492 266.744 0.023 0.002 False
7 9.002 334.877 17102.308 71152.492 266.744 0.023 0.002 False
8 9.002 334.877 17102.308 71152.492 266.744 0.023 0.002 False
9 9.002 334.877 17102.308 71152.492 266.744 0.023 0.002 False
10 9.002 334.877 17102.308 71152.492 266.744 0.023 0.002 False
11 9.002 334.877 17102.308 71152.492 266.744 0.023 0.002 False
12 9.002 334.877 17102.308 71152.492 266.744 0.023 0.002 False
13 9.002 334.877 17102.308 71152.492 266.744 0.023 0.002 False
14 9.002 334.877 17102.308 71152.492 266.744 0.023 0.002 False
15 9.002 334.877 17102.308 71152.492 266.744 0.023 0.002 False
16 9.002 334.877 17102.308 71152.492 266.744 0.023 0.002 False
17 9.002 334.877 17102.308 71152.492 266.744 0.023 0.002 False
18 9.002 334.877 17102.308 71152.492 266.744 0.023 0.002 False
19 9.002 334.877 17102.308 71152.492 266.744 0.023 0.002 False
20 9.002 334.877 17102.308 71152.492 266.744 0.023 0.002 False
21 9.002 334.877 17102.308 71152.492 266.744 0.023 0.002 False
22 9.002 334.877 17102.308 71152.492 266.744 0.023 0.002 False
23 9.002 334.877 17102.308 71152.492 266.744 0.023 0.002 False
24 9.002 334.877 17102.308 71152.492 266.744 0.023 0.002 False
25 9.002 334.877 17102.308 71152.492 266.744 0.023 0.002 False
26 9.002 334.877 17102.308 71152.492 266.744 0.023 0.002 False
27 9.002 334.877 17102.308 71152.492 266.744 0.023 0.002 False
28 9.002 334.877 17102.308 71152.492 266.744 0.023 0.002 False
29 9.002 334.877 17102.308 71152.492 266.744 0.023 0.002 False
30 9.002 334.877 17102.308 71152.492 266.744 0.023 0.002 False
31 9.002 334.877 17102.308 71152.492 266.744 0.023 0.002 False
32 9.002 334.877 17102.308 71152.492 266.744 0.023 0.002 False
33 9.002 334.877 17102.308 71152.492 266.744 0.023 0.002 False
34 9.002 334.877 17102.308 71152.492 266.744 0.023 0.002 False
35 9.002 334.877 17102.308 71152.492 266.744 0.023 0.002 False
36 9.002 334.877 17102.308 71152.492 266.744 0.023 0.002 False
37 9.002 334.877 17102.308 71152.492 266.744 0.023 0.002 False
38 9.002 334.877 17102.308 71152.492 266.744 0.023 0.002 False
39 9.002 334.877 17102.308 71152.492 266.744 0.023 0.002 False
40 9.002 334.877 17102.308 71152.492 266.744 0.023 0.002 False
41 9.002 334.877 17102.308 71152.492 266.744 0.023 0.002 False
42 9.002 334.877 17102.308 71152.492 266.744 0.023 0.002 False
43 9.002 334.877 17102.308 71152.492 266.744 0.023 0.002 False
44 9.002 334.877 17102.308 71152.492 266.744 0.023 0.002 False
45 9.002 334.877 17102.308 71152.492 266.744 0.023 0.002 False
46 9.002 334.877 17102.308 71152.492 266.744 0.023 0.002 False
47 9.002 334.877 17102.308 71152.492 266.744 0.023 0.002 False
48 9.002 334.877 17102.308 71152.492 266.744 0.023 0.002 False
49 9.002 334.877 17102.308 71152.492 266.744 0.023 0.002 False
50 9.002 334.877 17102.308 71152.492 266.744 0.023 0.002 False
51 9.002 334.877 17102.308 71152.492 266.744 0.023 0.002 False
52 9.002 334.877 17102.308 71152.492 266.744 0.023 0.002 False
53 9.002 334.877 17102.308 71152.492 266.744 0.023 0.002 False
54 9.002 334.877 17102.308 71152.492 266.744 0.023 0.002 False
55 9.002 334.877 17102.308 71152.492 266.744 0.023 0.002 False
56 9.002 334.877 17102.308 71152.492 266.744 0.023 0.002 False
57 9.002 334.877 17102.308 71152.492 266.744 0.023 0.002 False
58 9.002 334.877 17102.308 71152.492 266.744 0.023 0.002 False
59 8.777 330.280 17088.763 71039.832 266.533 0.023 0.002 False
60 22.919 372.735 17142.007 71483.197 267.363 0.027 0.002 False
61 29.354 396.004 17198.409 71954.376 268.243 0.027 0.002 False
62 33.833 410.837 17257.941 72453.369 269.172 0.028 0.002 False
63 37.289 422.112 17320.568 72980.176 270.148 0.028 0.002 False
64 40.103 430.901 17386.258 73534.796 271.173 0.028 0.002 False
65 42.446 438.336 17454.976 74117.229 272.245 0.028 0.002 False
66 44.485 444.569 17526.687 74727.477 273.363 0.029 0.002 False
67 46.228 450.113 17601.354 75365.538 274.528 0.029 0.002 False
68 47.856 454.952 17678.939 76031.412 275.738 0.029 0.002 False
69 49.232 459.388 17759.405 76725.100 276.993 0.030 0.002 False
70 50.469 463.365 17842.712 77446.602 278.292 0.030 0.002 False
71 51.660 467.085 17928.820 78195.918 279.635 0.030 0.002 False
72 52.752 470.487 18017.691 78973.047 281.021 0.031 0.002 False
73 53.746 473.715 18109.282 79777.989 282.450 0.031 0.002 False
74 54.647 476.713 18203.552 80610.746 283.920 0.031 0.002 False
75 55.447 479.589 18300.461 81471.315 285.432 0.032 0.002 False
76 56.158 482.292 18399.967 82359.699 286.984 0.031 0.003 False
77 56.771 484.908 18502.028 83275.896 288.576 0.031 0.003 False
78 57.284 487.391 18606.601 84219.907 290.207 0.030 0.003 False
79 57.740 489.809 18713.646 85191.731 291.876 0.029 0.003 False
80 58.207 492.124 18823.118 86191.369 293.584 0.029 0.003 False
81 58.662 494.391 18934.977 87218.821 295.328 0.028 0.003 False
82 59.098 496.575 19049.180 88274.086 297.110 0.028 0.003 False
83 59.517 498.723 19165.686 89357.165 298.927 0.027 0.003 False
84 59.918 500.805 19284.452 90468.057 300.779 0.027 0.003 False
85 60.301 502.859 19405.438 91606.763 302.666 0.026 0.003 False
86 60.665 504.861 19528.601 92773.283 304.587 0.026 0.003 False
87 61.008 506.839 19653.902 93967.616 306.541 0.025 0.003 False
88 61.324 508.776 19781.299 95189.763 308.528 0.025 0.003 False
89 61.589 510.693 19910.752 96439.724 310.547 0.025 0.003 False
90 61.774 512.577 20042.221 97717.498 312.598 0.024 0.003 False
91 61.870 514.444 20175.667 99023.086 314.679 0.024 0.003 False
92 61.905 516.284 20311.051 100356.487 316.791 0.024 0.003 False
93 61.915 518.109 20448.335 101717.703 318.932 0.023 0.003 False
94 61.909 519.912 20587.480 103106.731 321.102 0.023 0.003 False
95 61.895 521.699 20728.449 104523.573 323.301 0.023 0.003 False
96 61.879 523.470 20871.205 105968.229 325.528 0.022 0.003 False
97 61.866 525.223 21015.711 107440.699 327.781 0.022 0.003 False
98 61.860 526.962 21161.932 108940.982 330.062 0.022 0.003 False
99 61.889 528.681 21309.833 110469.079 332.369 0.022 0.003 False
100 61.963 530.389 21459.378 112024.989 334.701 0.021 0.003 False
101 61.958 532.070 21610.534 113608.713 337.059 0.021 0.003 False
102 61.856 533.743 21763.267 115220.251 339.441 0.021 0.003 False
103 61.688 535.382 21917.544 116859.602 341.847 0.021 0.003 False
104 61.466 537.010 22073.332 118526.767 344.277 0.020 0.002 False
105 61.198 538.595 22230.600 120221.746 346.730 0.020 0.002 False
106 60.939 540.167 22389.317 121944.538 349.206 0.020 0.002 False
107 60.676 541.678 22549.453 123695.143 351.703 0.020 0.002 False
108 60.496 543.170 22710.976 125473.563 354.222 0.020 0.002 False
109 60.210 544.577 22873.858 127279.796 356.763 0.020 0.002 False
110 59.770 545.951 23038.069 129113.842 359.324 0.019 0.002 False
111 59.628 547.201 23203.583 130975.703 361.906 0.019 0.002 False
112 59.415 548.391 23370.370 132865.376 364.507 0.019 0.002 False
113 59.725 549.387 23538.404 134782.864 367.128 0.019 0.002 False
114 61.103 550.267 23707.659 136728.165 369.768 0.019 0.002 False
115 69.506 550.816 23878.108 138701.280 372.426 0.018 0.002 False
116 79.447 551.112 24049.726 140702.208 375.103 0.018 0.002 False
117 80.185 550.742 24222.488 142730.950 377.797 0.018 0.002 False
118 76.880 549.676 24396.371 144787.505 380.510 0.018 0.002 False
119 72.937 546.584 24571.349 146871.875 383.239 0.018 0.002 False
120 92.632 572.153 24747.400 148984.057 385.985 0.016 0.002 False
121 93.588 577.735 24760.291 149139.310 386.186 0.016 0.002 False
122 93.588 577.735 24760.291 149139.310 386.186 0.016 0.002 False
123 93.588 577.735 24760.291 149139.310 386.186 0.016 0.002 False
124 93.588 577.735 24760.291 149139.310 386.186 0.016 0.002 False
125 93.588 577.735 24760.291 149139.310 386.186 0.016 0.002 False
126 93.588 577.735 24760.291 149139.310 386.186 0.016 0.002 False
127 93.588 577.735 24760.291 149139.310 386.186 0.016 0.002 False
128 93.588 577.735 24760.291 149139.310 386.186 0.016 0.002 False
129 93.588 577.735 24760.291 149139.310 386.186 0.016 0.002 False
130 93.588 577.735 24760.291 149139.310 386.186 0.016 0.002 False
131 93.588 577.735 24760.291 149139.310 386.186 0.016 0.002 False
132 93.588 577.735 24760.291 149139.310 386.186 0.016 0.002 False
133 93.588 577.735 24760.291 149139.310 386.186 0.016 0.002 False
134 93.588 577.735 24760.291 149139.310 386.186 0.016 0.002 False
135 93.588 577.735 24760.291 149139.310 386.186 0.016 0.002 False
136 93.588 577.735 24760.291 149139.310 386.186 0.016 0.002 False
137 93.588 577.735 24760.291 149139.310 386.186 0.016 0.002 False
138 93.588 577.735 24760.291 149139.310 386.186 0.016 0.002 False
139 93.588 577.735 24760.291 149139.310 386.186 0.016 0.002 False
140 93.588 577.735 24760.291 149139.310 386.186 0.016 0.002 False
141 93.588 577.735 24760.291 149139.310 386.186 0.016 0.002 False
142 93.588 577.735 24760.291 149139.310 386.186 0.016 0.002 False
143 93.588 577.735 24760.291 149139.310 386.186 0.016 0.002 False
144 93.588 577.735 24760.291 149139.310 386.186 0.016 0.002 False
145 93.588 577.735 24760.291 149139.310 386.186 0.016 0.002 False
146 93.588 577.735 24760.291 149139.310 386.186 0.016 0.002 False
147 93.588 577.735 24760.291 149139.310 386.186 0.016 0.002 False
148 93.588 577.735 24760.291 149139.310 386.186 0.016 0.002 False
149 93.588 577.735 24760.291 149139.310 386.186 0.016 0.002 False
150 93.588 577.735 24760.291 149139.310 386.186 0.016 0.002 False
151 93.588 577.735 24760.291 149139.310 386.186 0.016 0.002 False
152 93.588 577.735 24760.291 149139.310 386.186 0.016 0.002 False
153 93.588 577.735 24760.291 149139.310 386.186 0.016 0.002 False
154 93.588 577.735 24760.291 149139.310 386.186 0.016 0.002 False
155 93.588 577.735 24760.291 149139.310 386.186 0.016 0.002 False
156 93.588 577.735 24760.291 149139.310 386.186 0.016 0.002 False
157 93.588 577.735 24760.291 149139.310 386.186 0.016 0.002 False
158 93.588 577.735 24760.291 149139.310 386.186 0.016 0.002 False
159 93.588 577.735 24760.291 149139.310 386.186 0.016 0.002 False
160 93.588 577.735 24760.291 149139.310 386.186 0.016 0.002 False
161 93.588 577.735 24760.291 149139.310 386.186 0.016 0.002 False
162 93.588 577.735 24760.291 149139.310 386.186 0.016 0.002 False
163 93.588 577.735 24760.291 149139.310 386.186 0.016 0.002 False
164 93.588 577.735 24760.291 149139.310 386.186 0.016 0.002 False
165 93.588 577.735 24760.291 149139.310 386.186 0.016 0.002 False
166 93.588 577.735 24760.291 149139.310 386.186 0.016 0.002 False
167 93.588 577.735 24760.291 149139.310 386.186 0.016 0.002 False
168 93.588 577.735 24760.291 149139.310 386.186 0.016 0.002 False
169 93.588 577.735 24760.291 149139.310 386.186 0.016 0.002 False
170 93.588 577.735 24760.291 149139.310 386.186 0.016 0.002 False
171 93.588 577.735 24760.291 149139.310 386.186 0.016 0.002 False
172 93.588 577.735 24760.291 149139.310 386.186 0.016 0.002 False
173 93.588 577.735 24760.291 149139.310 386.186 0.016 0.002 False
174 93.588 577.735 24760.291 149139.310 386.186 0.016 0.002 False
175 93.588 577.735 24760.291 149139.310 386.186 0.016 0.002 False
176 93.588 577.735 24760.291 149139.310 386.186 0.016 0.002 False
177 93.588 577.735 24760.291 149139.310 386.186 0.016 0.002 False
178 93.588 577.735 24760.291 149139.310 386.186 0.016 0.002 False
179 93.588 577.735 24760.291 149139.310 386.186 0.016 0.002 False

}\loadedtable

\begin{table}[]
    \centering
\begin{tabular}{|r|l|r|r|r|r|r|} \hline
    $n, d$&$\sigma$& $MQH_{\sigma}$ & $MQHE_{\sigma}$ & $MCHE_{\sigma}  $
    &  $MSH_{\sigma}$&$MH_{\sigma}$\\ \hline    
20, 4  & stationary &3.07& 119.96& 323.44& 4.93& 24.30\\
&random& 3.37& 125.00& 325.12& 4.93& 24.34\\
&Dirac-like& 7.63& 177.79& 313.31& 4.87& 24.34\\
 \hline
40, 5  & stationary &4.45& 142.22& 450.37& 6.95& 48.29\\
&random& 4.82& 147.31& 452.92& 6.95& 48.29\\ 
&Dirac-like&8.45& 190.09& 444.83& 6.93& 48.29 \\
\hline
50, 6 & stationary &4.90& 148.85& 492.08& 7.61& 57.94\\
&random& 5.25& 153.36& 491.52& 7.62& 58.03 \\ 
&Dirac-like& 8.42& 190.51& 488.15& 7.60& 58.03 \\ \hline
100, 8 & stationary &6.91& 175.54& 682.39& 10.61& 112.55\\
&random& 7.37& 180.39& 682.23& 10.61&112.55 \\ 
&Dirac-like& 9.57& 204.40& 680.35& 10.60&112.55 \\ \hline
    \end{tabular}
    \caption{Random $d$-regular graphs. We report the average results over 10 randomly generated graphs. We notice that mean hitting time is independent of the initial distribution while quantum hitting time depends on the initial distribution.  }
    \label{tab:RR}
\end{table}

\pgfplotstableread{
node qh upq ubh ht sht e1 e1 tr

0 0.490 124.224 482.459 56.542 7.519 0.524 0.094 True
1 3.570 129.916 657.794 105.107 10.252 0.436 0.094 True
2 3.572 129.921 658.836 105.440 10.268 0.436 0.094 True
3 9.973 209.599 676.334 111.116 10.541 0.186 0.094 True
4 10.262 212.446 686.182 114.375 10.695 0.180 0.094 True
5 10.393 213.693 692.432 116.468 10.792 0.178 0.093 True
6 10.142 211.288 683.808 113.585 10.658 0.183 0.093 True
7 10.188 211.731 686.194 114.379 10.695 0.182 0.093 True
8 10.241 212.230 688.129 115.025 10.725 0.181 0.093 True
9 10.109 210.920 679.754 112.242 10.594 0.183 0.094 True
10 10.143 211.254 682.852 113.268 10.643 0.182 0.094 True
11 10.184 211.678 681.498 112.819 10.622 0.182 0.094 True
12 10.097 210.838 680.968 112.643 10.613 0.183 0.094 True
13 10.255 212.342 684.288 113.744 10.665 0.180 0.094 True
14 10.246 212.272 684.532 113.826 10.669 0.181 0.094 True
15 10.264 212.433 684.738 113.894 10.672 0.180 0.094 True
16 10.274 212.522 685.268 114.070 10.680 0.180 0.094 True
17 10.322 213.035 689.704 115.552 10.750 0.179 0.093 True
18 10.259 212.386 684.984 113.976 10.676 0.180 0.094 True
19 10.102 210.847 679.884 112.285 10.596 0.183 0.094 True
20 10.189 211.725 681.891 112.949 10.628 0.182 0.094 True
21 9.993 209.802 678.356 111.781 10.573 0.185 0.094 True
22 10.107 210.905 680.960 112.641 10.613 0.183 0.094 True
23 10.100 210.852 680.406 112.457 10.605 0.183 0.094 True
24 10.255 212.369 685.481 114.141 10.684 0.180 0.094 True
25 10.133 211.135 680.678 112.547 10.609 0.183 0.094 True
26 10.187 211.708 686.637 114.527 10.702 0.182 0.093 True
27 10.121 211.046 681.460 112.806 10.621 0.183 0.094 True
28 10.196 211.772 681.235 112.732 10.618 0.181 0.094 True
29 10.307 212.850 687.322 114.755 10.712 0.179 0.094 True
30 10.087 210.721 684.507 113.817 10.669 0.184 0.093 True
31 10.193 211.777 682.670 113.207 10.640 0.182 0.094 True
32 10.121 211.025 680.058 112.342 10.599 0.183 0.094 True
33 10.264 212.435 685.109 114.017 10.678 0.180 0.094 True
34 10.358 213.360 690.570 115.842 10.763 0.179 0.093 True
35 10.288 212.651 685.601 114.181 10.686 0.180 0.094 True
36 10.248 212.261 683.384 113.444 10.651 0.181 0.094 True
37 10.239 212.200 683.744 113.564 10.657 0.181 0.094 True
38 3.555 129.489 662.178 106.513 10.321 0.442 0.094 True
39 10.158 211.392 682.584 113.179 10.639 0.182 0.094 True
40 10.160 211.451 684.329 113.758 10.666 0.182 0.093 True
41 10.104 210.900 681.240 112.733 10.618 0.183 0.094 True
42 10.301 212.808 687.298 114.747 10.712 0.180 0.094 True
43 3.572 129.932 660.049 105.829 10.287 0.436 0.093 True
44 10.326 213.041 688.675 115.208 10.733 0.179 0.094 True
45 10.227 212.065 682.533 113.162 10.638 0.181 0.094 True
46 10.197 211.804 682.627 113.193 10.639 0.182 0.094 True
47 10.190 211.730 681.735 112.897 10.625 0.182 0.094 True
48 10.242 212.230 684.374 113.773 10.666 0.181 0.094 True
49 9.903 208.915 675.954 110.991 10.535 0.187 0.094 True
50 3.562 129.633 660.040 105.826 10.287 0.440 0.094 True
51 10.104 210.876 680.601 112.522 10.608 0.183 0.094 True
52 10.104 210.887 682.155 113.036 10.632 0.183 0.094 True
53 10.128 211.117 681.457 112.805 10.621 0.183 0.094 True
54 10.255 212.325 683.826 113.591 10.658 0.180 0.094 True
55 10.250 212.282 683.743 113.563 10.657 0.181 0.094 True
56 10.213 211.962 683.409 113.452 10.651 0.181 0.094 True
57 10.220 212.003 682.594 113.182 10.639 0.181 0.094 True
58 10.239 212.185 683.421 113.456 10.652 0.181 0.094 True
59 10.335 213.122 688.808 115.252 10.736 0.179 0.094 True
60 10.139 211.222 683.551 113.500 10.654 0.183 0.094 True
61 10.216 211.958 681.999 112.985 10.629 0.181 0.094 True
62 10.257 212.377 684.981 113.975 10.676 0.180 0.094 True
63 10.283 212.630 687.034 114.659 10.708 0.180 0.094 True
64 10.142 211.244 681.800 112.919 10.626 0.182 0.094 True
65 10.201 211.810 686.255 114.399 10.696 0.181 0.094 True
66 10.028 210.140 680.132 112.367 10.600 0.185 0.094 True
67 10.311 212.900 687.733 114.893 10.719 0.179 0.094 True
68 10.261 212.381 684.088 113.678 10.662 0.180 0.094 True
69 10.187 211.691 686.673 114.539 10.702 0.182 0.093 True
70 10.123 211.078 682.190 113.048 10.632 0.183 0.094 True
71 3.573 129.952 660.587 106.002 10.296 0.436 0.093 True
72 10.361 213.377 690.421 115.793 10.761 0.179 0.093 True
73 10.016 210.022 678.488 111.824 10.575 0.185 0.094 True
74 10.130 211.136 683.734 113.560 10.656 0.183 0.094 True
75 10.164 211.450 684.435 113.793 10.667 0.182 0.094 True
76 10.341 213.184 689.274 115.408 10.743 0.179 0.093 True
77 10.097 210.829 680.887 112.617 10.612 0.183 0.094 True
78 10.246 212.249 683.707 113.551 10.656 0.181 0.094 True
79 10.099 210.840 681.329 112.763 10.619 0.183 0.094 True
80 9.971 209.586 676.779 111.262 10.548 0.186 0.094 True
81 10.299 212.771 687.157 114.700 10.710 0.180 0.094 True
82 10.096 210.801 680.595 112.520 10.608 0.183 0.094 True
83 10.320 212.996 688.716 115.221 10.734 0.179 0.093 True
84 10.175 211.605 686.176 114.373 10.695 0.182 0.093 True
85 10.169 211.528 686.026 114.323 10.692 0.182 0.093 True
86 10.348 213.262 690.061 115.672 10.755 0.179 0.093 True
87 10.108 210.912 680.092 112.354 10.600 0.183 0.094 True
88 10.329 213.049 687.824 114.923 10.720 0.179 0.094 True
89 10.298 212.761 686.404 114.449 10.698 0.180 0.094 True
90 10.267 212.468 685.046 113.997 10.677 0.180 0.094 True
91 10.102 210.858 680.085 112.352 10.600 0.183 0.094 True
92 10.270 212.500 685.407 114.117 10.683 0.180 0.094 True
93 10.073 210.566 681.555 112.838 10.623 0.184 0.094 True
94 10.280 212.595 685.996 114.313 10.692 0.180 0.094 True
95 10.172 211.544 684.217 113.721 10.664 0.182 0.094 True
96 10.204 211.881 682.925 113.292 10.644 0.181 0.094 True
97 10.159 211.402 682.726 113.226 10.641 0.182 0.094 True
98 3.560 129.523 667.187 108.130 10.399 0.442 0.093 True
99 3.554 129.469 657.957 105.159 10.255 0.442 0.094 True
}\loadedtable

\section{Conclusions}\label{sec:conclusions}

In this work we have proposed a theoretical and experimental analysis of notions of quantum hitting time, compared to their classical counterparts. The main focus is on the speedup of quantum versus classical hitting time, implying faster diffusion and computational improvements in search algorithms;  but it is also interesting to note how the quantum walk framework can help highlight properties of graphs that random walks might not detect. As an example, consider the use of hitting time as a centrality measure for ranking purposes: for a barbell graph and a choice of stationary distribution (Figure \ref{fig:barbell-pi}) the quantum ranking is able to distinguish among nodes in the ``bar'' much more clearly than the classical ranking.  

Defining and studying quantum hitting time w.r.t.~ a general distribution presents several challenges. In particular, it is not easy to predict whether quadratic speedup will hold; see the discussion at the end of Section \ref{sec:generalized}. As further work, we would like to develop analytically bounds for quantum hitting time in special cases. For instance, note that in the time-reversible case the matrix $P_M$ is similar to a symmetric matrix and therefore has real eigenvalues and eigenvectors; it is possible to bound the right-hand side of \eqref{eq: HTsigma} under this hypothesis? Moreover, experimental evidence suggests that the first term $x_1y_1$ in \eqref{eq: HTsigma} is strongly dominant: it would be interesting to understand whether this property is true in general and how it can help formulate bounds for \eqref{eq: HTsigma}.

One driving idea behind the design of the experiments proposed here is to understand up to what point quantum hitting time depends on the associated distribution. The experimental evidence obtained so far suggests that the dependence is rather marked, in contrast, for instance, with behavior of classical hitting time for random graphs.  
Also note that the experimental part of the present work is focused on the case where there is a single marked vertex in the graph. Further developments will include extensive numerical experiments with $m>1$. 

\bibliographystyle{plain}
\bibliography{qwalksbib}

\begin{thebibliography}{10}

\bibitem{aldous2002reversible}
David Aldous and James Fill.
\newblock Reversible markov chains and random walks on graphs, 2002.

\bibitem{altafini2022centrality}
D~Altafini, DA~Bini, V~Cutini, B~Meini, and F~Poloni.
\newblock A centrality score of graph edges based on the kemeny constant.
\newblock {\em arXiv preprint arXiv:2203.06459}, 2022.

\bibitem{apers2022quadratic}
Simon Apers, Shantanav Chakraborty, Leonardo Novo, and J{\'e}r{\'e}mie Roland.
\newblock Quadratic speedup for spatial search by continuous-time quantum walk.
\newblock {\em Physical review letters}, 129(16):160502, 2022.

\bibitem{apers2022elfs}
Simon Apers and Stephen Piddock.
\newblock Elfs, trees and quantum walks.
\newblock {\em arXiv preprint arXiv:2211.16379}, 2022.

\bibitem{Bai2019}
Lu~Bai, Luca Rossi, Lixin Cui, Jian Cheng, and Edwin~R Hancock.
\newblock A quantum-inspired similarity measure for the analysis of complete
  weighted graphs.
\newblock {\em IEEE transactions on cybernetics}, 50(3):1264--1277, 2019.

\bibitem{bai2019quantum}
Lu~Bai, Luca Rossi, Lixin Cui, Jian Cheng, and Edwin~R Hancock.
\newblock A quantum-inspired similarity measure for the analysis of complete
  weighted graphs.
\newblock {\em IEEE transactions on cybernetics}, 50(3):1264--1277, 2019.

\bibitem{bini2018kemeny}
Dario Bini, Jeffrey~J Hunter, Guy Latouche, Beatrice Meini, and Peter Taylor.
\newblock Why is kemeny’s constant a constant?
\newblock {\em Journal of Applied Probability}, 55(4):1025--1036, 2018.

\bibitem{chakraborty2020finding}
Shantanav Chakraborty, Leonardo Novo, and J{\'e}r{\'e}mie Roland.
\newblock Finding a marked node on any graph via continuous-time quantum walks.
\newblock {\em Physical Review A}, 102(2):022227, 2020.

\bibitem{chen2008clustering}
Mo~Chen, Jianzhuang Liu, and Xiaoou Tang.
\newblock Clustering via random walk hitting time on directed graphs.
\newblock In {\em AAAI}, volume~8, pages 616--621, 2008.

\bibitem{ellens2011effective}
Wendy Ellens, Floske~M Spieksma, Piet Van~Mieghem, Almerima Jamakovic, and
  Robert~E Kooij.
\newblock Effective graph resistance.
\newblock {\em Linear algebra and its applications}, 435(10):2491--2506, 2011.

\bibitem{Emms2008}
David Emms, Edwin Hancock, and Richard Wilson.
\newblock Graph drawing using quantum commute time.
\newblock In {\em 2008 19th International Conference on Pattern Recognition},
  pages 1--4. IEEE, 2008.

\bibitem{estrada2010vibrational}
Ernesto Estrada and Naomichi Hatano.
\newblock A vibrational approach to node centrality and vulnerability in
  complex networks.
\newblock {\em Physica A: Statistical Mechanics and its Applications},
  389(17):3648--3660, 2010.

\bibitem{fasino2021hitting}
Dario Fasino, Arianna Tonetto, and Francesco Tudisco.
\newblock Hitting times for non-backtracking random walks.
\newblock {\em arXiv preprint arXiv:2105.14438}, 2021.

\bibitem{ghosh2008minimizing}
Arpita Ghosh, Stephen Boyd, and Amin Saberi.
\newblock Minimizing effective resistance of a graph.
\newblock {\em SIAM review}, 50(1):37--66, 2008.

\bibitem{SciPyProceedings_11}
Aric~A. Hagberg, Daniel~A. Schult, and Pieter~J. Swart.
\newblock Exploring network structure, dynamics, and function using networkx.
\newblock In Ga\"el Varoquaux, Travis Vaught, and Jarrod Millman, editors, {\em
  Proceedings of the 7th Python in Science Conference}, pages 11 -- 15,
  Pasadena, CA USA, 2008.

\bibitem{kadian2021quantum}
Karuna Kadian, Sunita Garhwal, and Ajay Kumar.
\newblock Quantum walk and its application domains: A systematic review.
\newblock {\em Computer Science Review}, 41:100419, 2021.

\bibitem{kempe2003proceedings}
J~Kempe.
\newblock Proceedings of the 7th international workshop on randomization and
  approximation techniques in computer science (random’03).
\newblock 2003.

\bibitem{kirkland2021directed}
Steve Kirkland.
\newblock Directed forests and the constancy of kemeny’s constant.
\newblock {\em Journal of Algebraic Combinatorics}, 53(1):81--84, 2021.

\bibitem{klein2010centrality}
DJ~Klein.
\newblock Centrality measure in graphs.
\newblock {\em Journal of mathematical chemistry}, 47(4):1209--1223, 2010.

\bibitem{krovi2006hitting}
Hari Krovi and Todd~A Brun.
\newblock Hitting time for quantum walks on the hypercube.
\newblock {\em Physical Review A}, 73(3):032341, 2006.

\bibitem{krovi2006quantum}
Hari Krovi and Todd~A Brun.
\newblock Quantum walks with infinite hitting times.
\newblock {\em Physical Review A}, 74(4):042334, 2006.

\bibitem{krovi2016quantum}
Hari Krovi, Fr{\'e}d{\'e}ric Magniez, Maris Ozols, and J{\'e}r{\'e}mie Roland.
\newblock Quantum walks can find a marked element on any graph.
\newblock {\em Algorithmica}, 74(2):851--907, 2016.

\bibitem{krovi2010adiabatic}
Hari Krovi, Maris Ozols, and J{\'e}r{\'e}mie Roland.
\newblock Adiabatic condition and the quantum hitting time of markov chains.
\newblock {\em Physical Review A}, 82(2):022333, 2010.

\bibitem{liben2003link}
David Liben-Nowell and Jon Kleinberg.
\newblock The link prediction problem for social networks.
\newblock In {\em Proceedings of the twelfth international conference on
  Information and knowledge management}, pages 556--559, 2003.

\bibitem{magniez2009proceedings}
Fr{\'e}d{\'e}ric Magniez, Ashwin Nayak, Peter~C Richter, and Miklos Santha.
\newblock Proceedings of the twentieth annual acm-siam symposium on discrete
  algorithms, soda'09.
\newblock 2009.

\bibitem{magniez2012hitting}
Fr{\'e}d{\'e}ric Magniez, Ashwin Nayak, Peter~C Richter, and Miklos Santha.
\newblock On the hitting times of quantum versus random walks.
\newblock {\em Algorithmica}, 63(1):91--116, 2012.

\bibitem{magniez2007search}
Fr{\'e}d{\'e}ric Magniez, Ashwin Nayak, J{\'e}r{\'e}mie Roland, and Miklos
  Santha.
\newblock Search via quantum walk.
\newblock In {\em Proceedings of the thirty-ninth annual ACM symposium on
  Theory of computing}, pages 575--584, 2007.

\bibitem{magniez2011search}
Fr{\'e}d{\'e}ric Magniez, Ashwin Nayak, J{\'e}r{\'e}mie Roland, and Miklos
  Santha.
\newblock Search via quantum walk.
\newblock {\em SIAM Journal on Computing}, 40(1):142--164, 2011.

\bibitem{palacios2010kirchhoff}
Jos{\'e}~Luis Palacios.
\newblock On the kirchhoff index of regular graphs.
\newblock {\em International Journal of Quantum Chemistry}, 110(7):1307--1309,
  2010.

\bibitem{paparo2014quantum}
Giuseppe~Davide Paparo, Markus M{\"u}ller, F~Comellas, and Miguel~Angel
  Martin-Delgado.
\newblock Quantum google algorithm.
\newblock {\em The European Physical Journal Plus}, 129(7):1--16, 2014.

\bibitem{paparo2013quantum}
Giuseppe~Davide Paparo, Markus M{\"u}ller, Francesc Comellas, and Miguel~Angel
  Martin-Delgado.
\newblock Quantum google in a complex network.
\newblock {\em Scientific reports}, 3(1):1--16, 2013.

\bibitem{PortugalBook}
Renato Portugal.
\newblock {\em Quantum walks and search algorithms}.
\newblock Springer, 2013.

\bibitem{roland2018finding}
J{\'e}r{\'e}mie Roland.
\newblock Finding a marked node on any graph by continuous-time quantum walk.
\newblock {\em arXiv preprint arXiv:1807.05957}, 2018.

\bibitem{Szegedy2004}
Mario Szegedy.
\newblock Quantum speed-up of {M}arkov chain based algorithms.
\newblock In {\em 45th Annual IEEE Symposium on Foundations of Computer
  science}, pages 32--41. IEEE, 2004.

\bibitem{szegedy2004quantum}
Mario Szegedy.
\newblock Quantum speed-up of markov chain based algorithms.
\newblock In {\em 45th Annual IEEE Symposium on Foundations of Computer
  Science}, pages 32--41. IEEE, 2004.

\bibitem{Varbanov2008}
Martin Varbanov, Hari Krovi, and Todd~A Brun.
\newblock Hitting time for the continuous quantum walk.
\newblock {\em Physical Review A}, 78(2):022324, 2008.

\bibitem{venegas2012quantum}
Salvador~El{\'\i}as Venegas-Andraca.
\newblock Quantum walks: a comprehensive review.
\newblock {\em Quantum Information Processing}, 11(5):1015--1106, 2012.

\bibitem{wang2017kemeny}
Xiangrong Wang, Johan~LA Dubbeldam, and Piet Van~Mieghem.
\newblock c.
\newblock {\em Linear Algebra and its Applications}, 535:231--244, 2017.

\bibitem{xia2019random}
Feng Xia, Jiaying Liu, Hansong Nie, Yonghao Fu, Liangtian Wan, and Xiangjie
  Kong.
\newblock Random walks: A review of algorithms and applications.
\newblock {\em IEEE Transactions on Emerging Topics in Computational
  Intelligence}, 4(2):95--107, 2019.

\end{thebibliography}

\end{document}